\documentclass[aps,showpacs,twocolumn,superscriptaddress]{revtex4-2}
\usepackage[utf8]{inputenc}
\usepackage{amsfonts}
\usepackage{amssymb}
\usepackage{amsmath}
\usepackage{graphicx}
\usepackage{epsfig}
\usepackage{subfigure}
\usepackage{appendix}
\usepackage{hyperref}
\usepackage{color}

\hypersetup{hypertex=true, colorlinks=true, linkcolor=blue, urlcolor=blue, citecolor=blue}

\begin{document}

\title{Dynamically stable topological edge states in an extended
Su-Schrieffer-Heeger ladder with balanced perturbation}
\author{E. S. Ma}
\affiliation{School of Physics, Nankai University, Tianjin 300071, China}
\author{K. L. Zhang}
\email{zhangkl@fosu.edu.cn}
\affiliation{School of Physics and Optoelectronic Engineering, Foshan University, Foshan 528225, China}
\author{Z. Song}
\email{songtc@nankai.edu.cn}
\affiliation{School of Physics, Nankai University, Tianjin 300071, China}
\begin{abstract}
The on-site potentials may break the symmetry of a system, resulting in the
loss of its original topology protected by the symmetry. In this work, we
study the counteracting effect of non-Hermitian terms on real potentials,
resulting in dynamically stable topological edge states. We show exactly for
a class of systems that the spectrum remains unchanged in the presence of
balanced perturbations. As a demonstration, we investigate an extended
non-Hermitian Su-Schrieffer-Heeger (SSH) ladder. We find that the
bulk-boundary correspondence still holds, and the zero-energy edge states
become coalescing states. In comparison to the original SSH chain, such edge
states are robust not only against local perturbations but also in the time
domain. As a result, a trivial initial state can always evolve to a stable
edge state. Our results provide insights for the application of time-domain
stable topological quantum devices.
\end{abstract}

\maketitle

\section{Introduction}

\label{Introduction}

Topological phases are characterized by the global properties of a system,
with one of their central ingredients being the topological edge states
protected by topology and symmetry of the system \cite{klitzing1980,
Thouless1982, Zak1989, Kitaev2001, Fu2007a, Nayak2008a, qi2011topological,
matsuura2013protected, Chiu2014, Sarma2015, Chiu2016, Asboth2016,
kobayashi2018symmetry, xie2023antihelical, rhim2017bulk, guzman2022geometry,
verma2024bulk}. Due to their resilience against local perturbations,
topological edge states hold significant potential for quantum computing and
highly robust devices \cite{Kitaev2001, Nayak2008a, Sarma2015, Else2017}.
Nowadays, non-equilibrium topological phases and the dynamic manipulation of
topological edge states have attracted great interest \cite%
{rechtsman2013photonic, Lohse2018, sun2018, zhang2018a, Li2019,
barbarino2020preparing, rudner2020, zhang2020unified, citro2023thouless,
bin2023out, lane2024extended} due to the development of related experimental
techniques in platforms such as ultracold atoms and optics. These
advancements have enabled unprecedented control over quantum systems,
allowing researchers to realize Floquet topological phases \cite%
{rechtsman2013photonic, Li2019} and topological charge pumping \cite%
{Lohse2018, citro2023thouless} through different driving protocols. Notably,
these platforms also allow us to introduce controllable gain and loss to
achieve non-Hermitian topological Hamiltonians.

Non-Hermiticity not only expands the scope of topological phases research
but also provides novel concepts and technical support for practical
applications in fields such as optics and acoustics \cite{Jin2017,
Takata2018, Harari2018, Bandres2018, Yao2018a, Yao2018, Jin2019, Lee2019a,
Luo2019, zhang2020non, zhou2021non, Wu2021, roccati2024hermitian,
huang2024acoustic}. On the one hand, the asymmetric hoppings and gain-loss
mechanisms in non-Hermitian systems offer additional degrees of freedom for
regulating topological properties, enabling the existence of exotic edge
states. Also, as the non-Hermitian skin effect (NHSE) may breaks the
bulk-boundary correspondence (BBC) and the original topological edge states,
new topological invariants are constructed to capture the topological
properties of such systems \cite{Yao2018, Jin2019, Lee2019a}. On the other
hand, non-Hermitian dynamics, such as exceptional point (EP) dynamics \cite%
{zhang2019exceptional, zhang2019helical, wang2021coherent,
wang2021transition, shi2022exceptional} and self-healing of non-Hermitian
skin modes \cite{longhi2022self, xue2025non, miao2025non}, provide a unique
pathway for flexibly preparing and controlling topological edge states
through external parameters. Recent studies have revealed that topological
skin-edge modes enabled by intrinsic non-Hermitian point-gap topology
exhibit a remarkable self-healing wave feature, demonstrating unprecedented
robustness against structural perturbations \cite{longhi2022self}. Based on
these considerations, an intriguing question is that in the case of line-gap
topology without NHSE, whether the non-Hermitian system support dynamically
stable topological edge states with self-healing characteristic?

In this work, we consider this question by constructing a class of lattice
model $\mathcal{H}=H_{0}+H_{\mathrm{p}}$, with Hermitian bipartite lattice $%
H_{0}$ and non-Hermitian perturbation term $H_{\mathrm{p}}$ consisting of
real on-site potentials and imaginary hoppings. The imaginary hoppings
connect the sites with the same index between the two sublattices and the
NHSE is absent. Thus, $\mathcal{H}$ exhibits line-gap spectrum. Under the
condition of balanced perturbations, we show exactly that the spectrum of $%
\mathcal{H}$ are the same as that of $H_{0}$. Notably, the zero modes of $%
H_{0}$ become coalescing states of $\mathcal{H}$, accompanied by the
appearance of EPs. Moreover, other eigenstates of the system can also turn
into coalescing states by tuning the real on-site potentials and imaginary
hoppings. As a demonstration, we investigate an extended non-Hermitian SSH
ladder model. The topological nature of this model can be characterized by
the sublattice winding number. We find that the BBC still holds, and with
balanced perturbations, the zero-energy edge states become coalescing states
with EP in the spectrum. In comparison to the original SSH chain, these edge
states are not only robust against local perturbations but also in the time
domain, exhibiting self-healing characteristic. Finally, we present
numerical results of time evolution to demonstrate that a trivial initial
state can always evolve to a dynamically stable topological edge state.

The rest of the paper is organized as follows. In Sec. \ref{Formalism}, we
introduce a class of lattice model and show that the eigenstates can be
turned into coalescing states on-demand by modulating the perturbations. In
Sec. \ref{Model}, we present an extended SSH model to demonstrate our
results, and in Sec. \ref{Dynamics}, we show that topological edge states
are stable in time domain. Finally, conclusions and discussions are given in
Sec. \ref{Conclusion}.

\section{Formalism}

\label{Formalism}

We consider a class of tight-binding model consists of a bipartite lattice $%
H_0$, non-Hermitian hoppings and on-site potentials $H_{\mathrm{p}}$, which
can be described by the Hamiltonian as follows%
\begin{equation}
\mathcal{H}=H_{0}+H_{\mathrm{p}},  \label{H}
\end{equation}%
with 
\begin{equation}
H_{0}=\sum_{i\neq j}J_{ij}a_{i}^{\dag }b_{j}+\mathrm{h.c.},
\end{equation}%
and 
\begin{equation}
H_{\mathrm{p}}=\sum\limits_{j=1}^{N}\left[ i\gamma\left( a_{j}^{\dag }b_{j}+%
\mathrm{h.c.}\right) +V\left( a_{j}^{\dag }a_{j}-b_{j}^{\dag }b_{j}\right) %
\right] .
\end{equation}%
This is a bipartite lattice system that can be decomposed into two
sublattices with equal site number $N_{\mathrm{a}}=N_{\mathrm{b}}=N$, where $%
a_{i}^{\dag }$ and $b_{j}^{\dag }$ ($a_{i}$ and $b_{j}$) denote the
creation (annihilation) operators on the $\mathrm{a}$ and $\mathrm{b}$ lattices,
and $J_{ij}$ is the hopping strength between the $\mathrm{a}$ lattice at $i$th
site and the $\mathrm{b}$ lattice at $j$th site with the constraint 
\begin{equation}
J_{ij}=-J_{ji},  \label{hopping_t}
\end{equation}
which is crucial for the following discussion. $\pm V$ is the on-site
potential for the $a$ and $b$ lattices, respectively. The imaginary hopping
strength between the $a$ and $b$ with the same site indexes is measured by $%
\gamma $.

In the following, we show that the diagonalization of $\mathcal{H}$ can be
resolved from that of $H_{0}$. In fact, $H_{0}$ and $\mathcal{H}$ can be
written in the following forms  
\begin{equation}
H_{0}=\Psi ^{\dag }\left( 
\begin{array}{cc}
\mathbf{0} & T \\ 
-T & \mathbf{0}%
\end{array}%
\right) \Psi ,
\end{equation}%
and%
\begin{equation}
\mathcal{H}=\Psi ^{\dag }\left( 
\begin{array}{cc}
VI_N & T+i\gamma I_N \\ 
-T+i\gamma I_N & -VI_N%
\end{array}%
\right) \Psi ,
\end{equation}%
where $\Psi ^{\dag }$ is defined as 
\begin{equation}
\Psi ^{\dag }=\left( a_{1}^{\dag },a_{2}^{\dag }\cdots a_{N}^{\dag
},b_{1}^{\dag },b_{2}^{\dag }\cdots b_{N}^{\dag }\right) ,
\end{equation}%
and $I_N$ is the $N\times N$ identity matrix, $T$ is an anti-symmetric $%
N\times N$ matrix with elements 
\begin{equation}
T_{ij}=t_{ij}.
\end{equation}

It can be checked that the diagonal form of $H_{0}$ is 
\begin{equation}
H_{0}=\sum\limits_{n=1}^{N}\sum_{\rho =\pm }\rho E_n D_{n,\rho }^{\dag
}D_{n,\rho },
\end{equation}%
where $D_{n,\rho }^{\dag }$ is defined as 
\begin{equation}
D_{n,\pm }^{\dag }=\frac{\Psi ^{\dag }}{\sqrt{2}}\left( 
\begin{array}{l}
\psi _{n} \\ 
\pm i\psi _{n}%
\end{array}%
\right) ,
\end{equation}%
with the $N\times 1$ column vector $\psi _{n}$ being the eigenstate of the
Hermitian matrix $iT$, that is 
\begin{equation}
iT\psi _{n}=E_n \psi _{n}.
\end{equation}%
Under the basis of $D_{n,\rho }^{\dag }$, the non-Hermitian Hamiltonian $%
\mathcal{H}$ can be rewritten as 
\begin{equation}
\mathcal{H} =\sum\limits_{n=1}^{N}\left( D_{j,+}^{\dag },D_{j,-}^{\dag
}\right) \left( 
\begin{array}{ll}
E_n & V+\gamma \\ 
V-\gamma & -E_n%
\end{array}%
\right) \left( 
\begin{array}{l}
D_{j,+} \\ 
D_{j,-}%
\end{array}%
\right) ,  \label{H_block}
\end{equation}
which can be further diagonalized as 
\begin{equation}
\mathcal{H}=\sum\limits_{n=1}^{N}\sum\limits_{\rho =\pm } \mathcal{E}%
_{n,\rho} \overline{\mathcal{D}}_{n,\rho }\mathcal{D}_{n,\rho },
\end{equation}
where 
\begin{equation}
\mathcal{E}_{n,\pm}=\pm\sqrt{E_n^2 +V^{2}-\gamma^{2}},  \label{spectrum}
\end{equation}%
and 
\begin{equation}
\begin{gathered}\overline{\mathcal{D}}_{n,\pm} =\frac{\left(
E_{n}+\mathcal{E}_{n,\pm} \right) D_{n,+}^{\dagger}+\left( V-\gamma \right)
D_{n,-}^{\dagger}}{\sqrt{(E_{n}+\mathcal{E}_{n,\pm} )^{2}+V^{2}-\gamma^{2}}}
,\\ \mathcal{D}_{n,\pm} =\frac{\left( E_{n}+\mathcal{E}_{n,\pm} \right)
D_{n,+}+\left( V+\gamma \right) D_{n,-}}{\sqrt{(E_{n}+\mathcal{E}_{n,\pm}
)^{2}+V^{2}-\gamma^{2}}}.\end{gathered}  \label{stateD}
\end{equation}
Therefore, the eigenstate of $\mathcal{H}$ with eigenenergy $\mathcal{E}%
_{n,\pm}$ is $\overline{\mathcal{D}}_{n,\pm}|0\rangle$, where $|0\rangle$ is
the vacuum state satisfying $a_j|0\rangle=b_j|0\rangle=0$. Note that $%
\overline{\mathcal{D}}_{n,\pm}\neq \mathcal{D}_{n,\pm}^{\dag}$ for the
non-Hermitian case $\gamma\neq 0$. Nevertheless, we note that the system
possesses a real spectrum or its eigenvalues always come in complex
conjugate pairs, since $\mathcal{H}$ has pseudo-Hermiticity \cite%
{Mostafazadeh2002a}, and the spectrum gap is line gap instead of point gap.

These results indicate that we can obtain the solutions of $\mathcal{H}$
via $H_{0}$, and it is worth noting that the conclusion does not depend
on the boundary conditions. From the solutions in Eqs. (\ref{spectrum}) and (%
\ref{stateD}), two conclusions can be drawn: (i) when $V=\pm \gamma$,
systems $\mathcal{H}$ and $H_{0}$ share the same spectrum, i.e., the
non-Hermitian hopping $i\gamma$ counteracts the effects of the on-site
potential $V$ from the perspective of energy; (ii) when $E_{n^{\prime }}=\pm 
\sqrt{\gamma^2-V^2}$, we have $\mathcal{E}_{n^{\prime },\pm}=0$ and $%
\overline{\mathcal{D}}_{n^{\prime },+}=\overline{\mathcal{D}}_{n^{\prime
},-} $, i.e., there are EPs in the spectrum, and two eigenstates coalesce into
the same one. In this case, the Hamiltonian in Eq. (\ref{H_block}) cannot be
diagonalized, and there will be peculiar dynamic behaviors. Considering the
time evolution for an initial state $(\alpha,\ \beta)_{n^{\prime }}^{T}$ in
the $n^{\prime }$th subspace, the evolved state is 
\begin{equation}
\begin{gathered}\exp \left[ -i\left( \begin{matrix}E_{n^{\prime}}&V+\gamma\\
V-\gamma&-E_{n^{\prime}}\end{matrix} \right)_{n'} t \right] \left(
\begin{gathered}\alpha\\ \beta\end{gathered} \right)_{n'}\\ =\left(
\begin{gathered}\alpha\\ \beta\end{gathered} \right)_{n'} -it\left(
\alpha+\frac{E_{n^{\prime}}}{\gamma -V} \beta \right) \left(
\begin{gathered}E_{n^{\prime}}\\ V-\gamma\end{gathered}
\right)_{n'}.\end{gathered}
\end{equation}
We can see that over time, the evolved state tends toward the coalescing
mode $\overline{\mathcal{D}}_{n^{\prime },+}|0\rangle$ as long as the
initial state has a nonzero overlap with state $(\alpha,\ \beta)_{n^{\prime
}}^{T}$ that satisfies $\alpha\neq E_{n^{\prime}}\beta/(V-\gamma)$. This is
called the EP dynamics.

Notably, the EPs are tunable by adjusting the parameters $\gamma$ and $%
V $, and at the same time, the on-site potential $V$ ensures the existence
of a full real spectrum. As a result, these EPs can be utilized to generate
stable quantum states on demand through the EP dynamics. Particularly, when $%
H_{0}$ is a topological system, this mechanism can be utilized to generate
dynamically stable topologically edge states. In the following, we apply
this formalism to an extended SSH ladder to demonstrate the above
conclusions.

\section{Extended SSH model}

\label{Model}

\subsection{Model and solution}

\begin{figure}[tbh]
\centering
\includegraphics[width=0.5\textwidth]{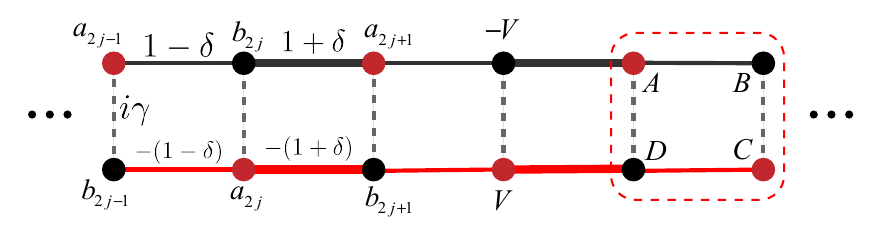}
\caption{The Schematic of the extended SSH ladder with imaginary tunneling.
The red (black) dots denote lattice sites $\mathrm{a}$ ($\mathrm{b}$), with
on-site potentials $V$ ($-V$). Each unit cell includes four sites denoted as
($A,B,C,D$).}
\label{fig1}
\end{figure}

In this section, we consider an extended SSH ladder with non-Hermitian
hoppings between the two legs of the ladder, as shown in Fig. \ref{fig1}. The
Hamiltonian has the same form as that in Eq. (\ref{H}) , which is%
\begin{equation}
\mathcal{H}=H_{0}+H_{\mathrm{p}},  \label{Hs}
\end{equation}%
where%
\begin{equation}
H_{0}=\sum_{j}J _{j,j+1}a_{j}^{\dag }b_{j+1}+\mathrm{h.c.},
\end{equation}%
and 
\begin{equation}
H_{\mathrm{p}}=\sum_{j}\left[ i\gamma \left( a_{j}^{\dag }b_{j}+\mathrm{h.c.}%
\right) +V\left( a_{j}^{\dag }a_{j}-b_{j}^{\dag }b_{j}\right) \right] .
\end{equation}%
Here the explicit expression of tunneling strength is 
\begin{equation}
J _{j,j+1}=-J _{j+1,j}=\left( -1\right) ^{j+1}\left[ 1+\left( -1\right)
^{j}\delta \right] ,
\end{equation}%
and the on-site potential on the $a$ lattice and $b$ lattice is $V$ and $-V$,
respectively. Here, $i\gamma $ represents the non-Hermitian hopping.

In the minimal topological model, the SSH chain \cite{Su1979}, the topological
properties of which arise from the alternating hopping amplitudes,
demonstrates a nontrivial topology when the winding number or Zak phase is
nonzero \cite{Zak1989, Asboth2016}. As a result, the boundaries of the
system host zero-energy edge states that are protected by the topology and
inversion symmetry of the model. For the extended SSH ladder we are considering,
the inversion symmetry is broken by the alternating on-site potential, as
in the Rice-Mele model \cite{rice1982elementary}. In the following, we
present the solution of the SSH ladder and show that it also supports a
topological phase characterized by the sublattice winding number even in the
presence of non-Hermitian terms.

Taking periodic boundary conditions for the system with $N$ unit cells, and
applying the Fourier transformation 
\begin{equation}
\left( 
\begin{array}{c}
a_{2j-1} \\ 
b_{2j} \\ 
a_{2j} \\ 
b_{2j-1}%
\end{array}%
\right) =\sum_{k}\frac{e^{ikj}}{\sqrt{N}}\left( 
\begin{array}{c}
a_{k,1} \\ 
b_{k,1} \\ 
a_{k,2} \\ 
b_{k,2}%
\end{array}%
\right) ,
\end{equation}%
where $k=2\pi n/N,\left( n=0,1,\cdots N-1\right) ,$ then the Hamiltonian can
be expressed as%
\begin{equation}
\mathcal{H}=\sum_{k}\Psi ^{\dag }h_{k}\Psi ,
\end{equation}%
Here, the operator vector is defined as%
\begin{equation}
\Psi ^{\dag }=\left( 
\begin{array}{cccc}
a_{k,1}^{\dag } & b_{k,1}^{\dag } & a_{k,2}^{\dag } & -b_{k,2}^{\dag }%
\end{array}%
\right) ,
\end{equation}%
and the Bloch Hamiltonian is%
\begin{equation}
h_{k}=\left( 
\begin{array}{cccc}
V & \xi_{k} & 0 & -i\gamma \\ 
\xi_{k} ^{\ast } & -V & i\gamma & 0 \\ 
0 & i\gamma & V & \xi_{k} ^{\ast } \\ 
-i\gamma & 0 & \xi_{k} & -V%
\end{array}%
\right) ,  \label{core}
\end{equation}%
where the $k$-dependent parameter is given by%
\begin{equation}
\xi_{k} =1-\delta +\left( 1+\delta \right) e^{-ik}.
\end{equation}

The Hamiltonian respects the following symmetry 
\begin{equation}
SKh_{k}\left( SK\right) ^{-1}=h_{k},
\end{equation}%
where 
\begin{equation}
S=\left( 
\begin{array}{cccc}
0 & 0 & 1 & 0 \\ 
0 & 0 & 0 & 1 \\ 
1 & 0 & 0 & 0 \\ 
0 & 1 & 0 & 0%
\end{array}%
\right) ,
\end{equation}%
and $K$ is the complex conjugate operator defined as $KiK=-i$. This means
that for an eigenstate $\left\vert \psi \right\rangle $ of $h_{k}$ with
eigenenergy $\varepsilon _{k},$ the state $SK\left\vert \psi \right\rangle $ is
also the eigenstate of $h_{k}$ with the same eigenvalue, i.e., the spectrum
is at least two-fold degenerate. Actually, the eigenstates and eigenenergies
of $h_{k}$ can be given as 
\begin{equation}
\left\vert \psi _{1}\right\rangle =\Omega_{1}\left( 
\begin{array}{c}
\xi _{k} \\ 
\varepsilon _{k}^{-}-V \\ 
i\gamma \\ 
0%
\end{array}%
\right) ,\left\vert \psi _{2}\right\rangle =\Omega_{2}\left( 
\begin{array}{c}
\varepsilon _{k}^{-}+V \\ 
\xi _{k}^{\ast } \\ 
0 \\ 
-i\gamma%
\end{array}%
\right) ,  
\end{equation}
with eigenenergy 
\begin{equation}
\varepsilon _{k}^{-}=-\sqrt{\xi _{k}\xi _{k}^{\ast }+V^{2}-\gamma ^{2}},
\end{equation}
and 
\begin{equation}
\left\vert \psi _{3}\right\rangle =\Omega_{3}\left( 
\begin{array}{c}
\xi _{k} \\ 
\varepsilon _{k}^{+}-V \\ 
i\gamma \\ 
0%
\end{array}%
\right) ,\left\vert \psi _{4}\right\rangle =\Omega_{4}\left( 
\begin{array}{c}
\varepsilon _{k}^{+}+V \\ 
\xi _{k}^{\ast } \\ 
0 \\ 
-i\gamma%
\end{array}%
\right) ,  
\end{equation}
with eigenenergy 
\begin{equation}
\varepsilon _{k}^{+}=\sqrt{\xi _{k}\xi _{k}^{\ast }+V^{2}-\gamma ^{2}}.
\end{equation}
Here $\Omega_{\alpha}$, $\alpha=1,2,3,4$, are the corresponding
normalization constants. Obviously, as discussed in the previous section,
when $\xi _{k^{\prime }}\xi _{k^{\prime }}^{\ast }=\gamma ^{2}-V^{2}$, there
are EPs in the spectrum, and $\varepsilon _{k^{\prime }}^{+}=\varepsilon
_{k^{\prime }}^{-}=0$. In this case, two pairs of degenerate eigenstates
coalesce into two eigenstates 
\begin{equation}
\left\vert \psi _{1}\right\rangle =\left\vert \psi _{3}\right\rangle,\
\left\vert \psi _{2}\right\rangle =\left\vert \psi _{4}\right\rangle .
\end{equation}

\begin{figure*}[tbh]
\centering
\includegraphics[width=1\textwidth]{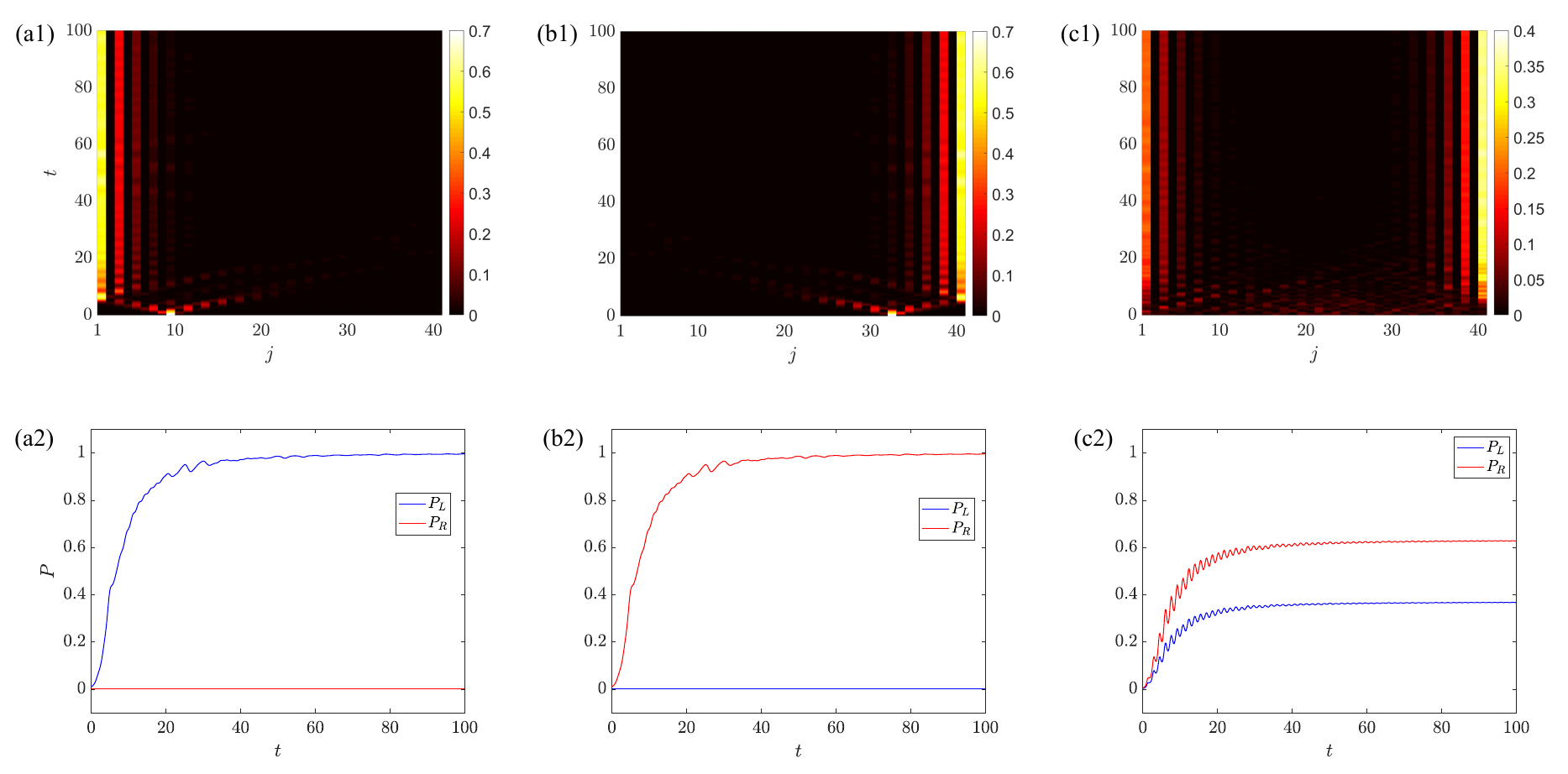}
\caption{The spatial distribution of the modulus of the amplitudes of the
evolved state (top panels) and the plots of the overlap between the evolved
state and two edge states as a function of time (bottom panels). For (a1)
and (b1), the initial states are two point excitations localized at the site
near the left end (the $9$th site) and the right end (the $32$nd site) of
the system, respectively. For (c1), the initial state is a state with a random
distribution. (a2), (b2) and (c2) give the corresponding overlap between the
evolved state and the left edge state ($P_{L}$) and the right edge state ($%
P_{R}$) as a function of time. Other parameters are taken as $N=20$, $V=\protect%
\gamma=1$, and $\protect\delta=0.2$. }
\label{fig2}
\end{figure*}

\subsection{Sublattice winding number}

To characterize the topology of the extended SSH ladder without inversion
symmetry and chiral symmetry, we employ the sublattice winding number \cite%
{rhim2017bulk, guzman2022geometry, verma2024bulk}, which is defined as 
\begin{equation}
w_{\epsilon }^{\rho }=\frac{1}{2\pi i}\oint\nolimits_{\mathrm{BZ}}dk\text{%
\textrm{Tr}}\left[ P_{\rho }\left( h_{k}-\epsilon \right) ^{-1}\partial
_{k}\left( h_{k}-\epsilon \right) P_{\rho }\right] ,  \label{subwinding}
\end{equation}%
where $\rho $ represents the sublattice index. For the SSH ladder shown in
Fig. \ref{fig1}, there are four sublattices in every unit cell, that is $%
\rho =A,B,C,D$, and $P_{\rho }=\left\vert \rho \right\rangle \left\langle
\rho \right\vert $ is the projection operator with 
\begin{equation}
\left\vert \rho \right\rangle =\left( \delta _{\rho ,A},\delta _{\rho
,B},\delta _{\rho ,C},\delta _{\rho ,D}\right) ^{T}.
\end{equation}%
Taking the reference energy as $\epsilon =\pm \sqrt{V^{2}-\gamma ^{2}}$,
direct derivations show that (see Appendix A) 
\begin{eqnarray}
w_{\epsilon }^{A} &=&\int_{0}^{2\pi }\frac{dk}{2\pi i}\partial _{k}\log
\left( \xi_k ^{\ast }\right) =\left\{ 
\begin{array}{c}
0,\ \delta <0 \\ 
1,\ \delta >0%
\end{array}%
\right. ,  \notag \\
w_{\epsilon }^{B} &=&\int_{0}^{2\pi }\frac{dk}{2\pi i}\partial _{k}\log
\left( \xi_k \right) =\left\{ 
\begin{array}{c}
0,\ \delta <0 \\ 
-1,\ \delta >0%
\end{array}%
\right. ,
\end{eqnarray}%
and 
\begin{equation}
w_{\epsilon }^{D}=w_{\epsilon }^{A},w_{\epsilon }^{C}=w_{\epsilon }^{B}.
\end{equation}

Furthermore, one can define a topological invariant 
\begin{equation}
w^{A}=\sum_{\epsilon }w_{\epsilon }^{A},
\end{equation}%
the nontrivial value of which equals the number of edge states distributed
at each end of the system. For the SSH ladder, the reference energy is
two-fold degenerate, therefore 
\begin{equation}
w^{A}=2w_{\epsilon }^{A}.
\end{equation}%
This indicates that there are $2$ edge states localized at each end of the
system. The topological edge states can be exactly solved and are dynamically
stable, as shown in the next section.

\section{Dynamically stable topological edge states}

\label{Dynamics}

Now we investigate the edge states and their dynamic behavior for the SSH
ladder. When taking the open boundary conditions, according to the
discussions in Sec. \ref{Formalism}, the eigenenergies of $\mathcal{H}$ are 
\begin{equation}
\mathcal{E} _{\pm }=\pm \sqrt{E_{0}^{2}+V^{2}-\gamma ^{2}},
\end{equation}%
where $E_{0}$ is the eigenenergy of $H_{0}$, and for the edge states of the
SSH chain, we have $E_{0}=0$. Thus, for the SSH ladder $\mathcal{H}$ , the
energies of the edge states are $\mathcal{E} _{\pm }=\pm \sqrt{V^{2}-\gamma
^{2}}$.

We set the ansatz of the edge state as the following form 
\begin{equation}
\left\vert \psi \right\rangle =\sum_{l=1}^{2N}\left( \alpha _{l}a_{l}^{\dag
}+\beta _{l}b_{l}^{\dag }\right) \left\vert 0\right\rangle ,
\end{equation}%
and substituting it and $\mathcal{H}$ into the Schr\"{o}dinger equation 
\begin{equation}
\mathcal{H}\left\vert \psi \right\rangle =\pm \sqrt{V^{2}-\gamma ^{2}}%
\left\vert \psi \right\rangle .  \label{Sch}
\end{equation}

Under the condition of $N\rightarrow \infty $ and $0<\delta<1$, by solving
the Schr\"{o}dinger equation in Eq. (\ref{Sch}), we obtain four edge states
(see Appendix B for the detials) 
\begin{eqnarray}
\left\vert \psi _{\mathrm{L}}^{\pm }\right\rangle &=&\Omega_{\mathrm{L}%
}^{\pm}\sum_{j=1}^{N}\left( -1\right) ^{j-1}e^{-\left( j-1\right) \zeta
}\left( \kappa _{\pm }a_{2j-1}^{\dag }+b_{2j-1}^{\dag }\right) \left\vert
0\right\rangle ,  \notag \\
\left\vert \psi _{\mathrm{R}}^{\pm }\right\rangle &=&\Omega_{\mathrm{R}%
}^{\pm}\sum_{j=1}^{N}\left( -1\right) ^{N-j}e^{-\left( N-j\right) \zeta
}\left( \kappa _{\pm }a_{2j}^{\dag }+b_{2j}^{\dag }\right) \left\vert
0\right\rangle ,  \label{edge_states}
\end{eqnarray}%
with $\Omega_{\mathrm{L}/\mathrm{R}}^{\pm}$ being the normalization
constant. Like the SSH chain, the amplitudes in Eq. (\ref{edge_states}) are
distributed at the lattice sites with odd (even) index for edge states
localized at the left (right) of the ladder. While the relative amplitudes
for lattices $\mathrm{a}$ and $\mathrm{b}$ at the same index are modulated
by $\kappa _{\pm }$, which is defined as 
\begin{equation}
\kappa _{\pm }=\frac{-i\gamma }{V\mp \sqrt{V^{2}-\gamma ^{2}}}.
\end{equation}
The amplitudes of edge states exhibit exponential decay, and the localization
length is $\zeta =\ln \left[ \left( \delta +1\right) /\left( 1-\delta
\right) \right] >0$, $\left( 0<\delta<1\right) $. The above results are
consistent with the results of the sublattice winding number obtained in the
previous section, establishing the BBC. Obviously, when $\left\vert
V\right\vert =\left\vert \lambda \right\vert $, the spectrum is fully real,
and there are EPs for the zero energy edge states. In this case 
\begin{equation}
\kappa _{+}=\kappa _{-},
\end{equation}%
thus, two pairs of edge states coalesce to two edge states 
\begin{equation}
\left\vert \psi _{\mathrm{L}}^{+}\right\rangle =\left\vert \psi _{\mathrm{L}%
}^{-}\right\rangle ,\left\vert \psi _{\mathrm{R}}^{+}\right\rangle
=\left\vert \psi _{\mathrm{R}}^{-}\right\rangle .
\end{equation}

The above two coalescing edge states are dynamically stable due to the EP
dynamics introduced in Sec. \ref{Formalism}. This can be verified by the
numerical computation of time evolution for the time-dependent Schr\"{o}%
dinger equation. Considering an arbitrary initial state in the following
form 
\begin{equation}
\left\vert \psi \left( 0\right) \right\rangle =\phi ^{\dag }\psi \left\vert
0 \right\rangle,
\end{equation}%
where 
\begin{equation}
\phi ^{\dag }=\left( a_{1}^{\dag },b_{2}^{\dag }\cdots b_{2N}^{\dag
},b_{1}^{\dag },a_{2}^{\dag }\cdots a_{2N}^{\dag }\right) ,
\end{equation}%
and $\psi $ is a $4N\times 1$ column vector. In the topological phase with
EP, one can expect that the initial state will converge to a stable edge
state composed of two coalescing states after a sufficiently long time. To
verify this prediction, numerical results for different initial states are
presented in Fig. \ref{fig2}. The normalized evolved state is $\left\vert
\psi \left( t \right) \right\rangle =e^{-i\mathcal{H} t}\left\vert \psi
\left( 0 \right) \right\rangle /\left| e^{-i\mathcal{H} t}\left\vert \psi
\left( 0 \right) \right\rangle \right|$, and the numerical computations are
performed by using a uniform mesh for time discretization. The top
panels of Fig. \ref{fig2} show the time evolution of the modulus of the
normalized amplitudes $\left| \langle j\left\vert \psi \left( t \right)
\right\rangle \right|^{2}$, where $\left\vert j \right\rangle =(
a_{j}^{\dag}+b_{j}^{\dag} ) \left\vert 0 \right\rangle$. The bottom panels
show the results of modulus of overlaps between the evolved state and the
normalized edge states in Eq. (\ref{edge_states}), that is $P_{L/R}=|
\langle \psi_{L/R}^{+} \left\vert \psi \left( t \right) \right\rangle|^{2}$.
Parameters of the system are taken as $N=20$, $V=\gamma=1$, and $\delta=0.2$.

Figs. \ref{fig2}(a) and (b) demonstrate that if the initial point excitation
localized at the left (right) of the system, then the stable final state
would converge to the left (right) edge state. In addition, Fig. \ref{fig2}%
(c) show that an initial state with random amplitude distribution would
evolve to an edge state distributed on the two ends of the system.

\begin{figure}[tbh]
\centering
\includegraphics[width=0.4\textwidth]{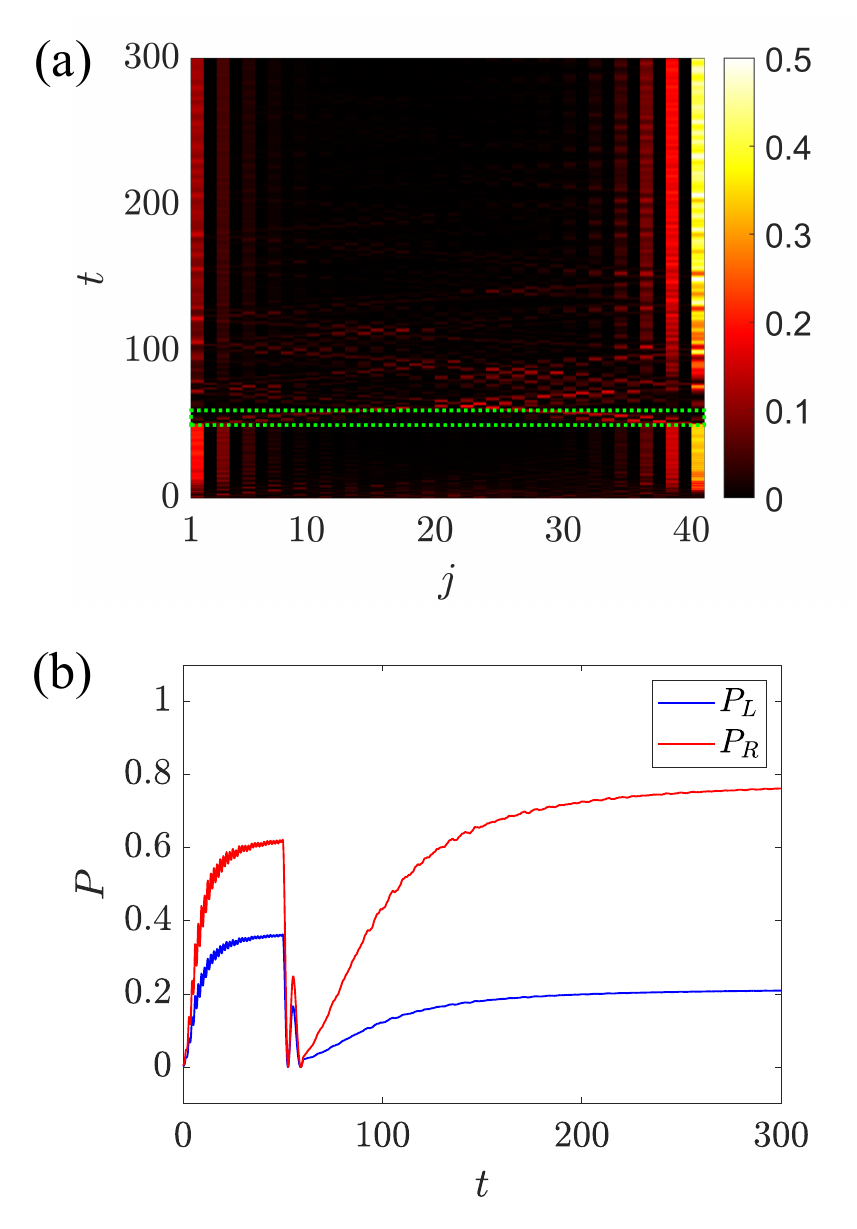}
\caption{(a) The spatial distribution of the modulus of the amplitudes of
the evolved state. The initial state is a state with a random distribution,
 the same as that in Fig. \protect\ref{fig2} (c1), and the parameter of
the initial Hamiltonian is $\protect\delta=0.2$. While there is a
perturbation in the temporal interval $t\in \left[ 50,60\right] $ marked by
the green dashed rectangle, in which the system is quenched to the parameter $%
\protect\delta=-0.2$. (b) The plot of the overlap between the evolved state
and two edge states as a function of time. Other parameters are taken as $N=20$ and $V=%
\protect\gamma=1$.}
\label{fig3}
\end{figure}

In order to investigate the robustness of the edge state in time domain, we
apply a global quench to parameter $\delta$ in a temporal interval. The
initial state and Hamiltonian are taken to be the same as that in Fig. \ref{fig2}
(c1). While at $t=50$, $\delta$ is switched from $0.2$ to $-0.2$, which corresponds to
the topologically trivial phase, and at $t=60$, $\delta$ is switched back to $%
0.2$. The numerical results of this quenching process are shown in Fig. \ref%
{fig3}. We can see that the formed edge state is disrupted in the temporal
interval $t\in \left[ 50,60\right] $, and gradually recovers after $t=60$.
These numerical results indicate that the zero-energy edge states in the SSH
ladder are robust in the time domain.

\section{Conclusion}

\label{Conclusion}

In summary, we have shown that in a class of non-Hermitian bipartite
lattices, the imaginary hopping and staggered on-site potential can be
utilized to generate tunable EP with two eigenstates coalescing into the same
one. We apply this mechanism to generate dynamically stable topological edge
states in an extended SSH ladder with balanced imaginary hopping and on-site
potential. In this model, the zero-energy edge states are characterized by
the sublattice winding number, and are not only robust against local
perturbations but also in the time domain. Numerical results demonstrate
that a trivial initial state can always evolve into a stable topological edge
state. Our results provide insights for the application of time-domain
stable topological quantum devices.

\acknowledgments This work was supported by the National Natural Science
Foundation of China (under Grant No. 12374461).

\section*{Appendix A: Calcultion of the sublattice winding number}

\label{A} \setcounter{equation}{0} \renewcommand{\theequation}{A%
\arabic{equation}}

In this appendix, we present the detailed calculation about the sublattice
winding number. According to Eq. (\ref{core}) we have%
\begin{equation}
\frac{1}{h_{k}-\epsilon }=\left( 
\begin{array}{cccc}
\frac{V+\epsilon }{\xi \xi ^{\ast }} & \frac{1}{\xi ^{\ast }} & 0 & \frac{%
-i\gamma }{\xi \xi ^{\ast }} \\ 
\frac{1}{\xi } & \frac{V-\epsilon }{-\xi \xi ^{\ast }} & \frac{i\gamma }{\xi
\xi ^{\ast }} & 0 \\ 
0 & \frac{i\gamma }{\xi \xi ^{\ast }} & \frac{V+\epsilon }{\xi \xi ^{\ast }}
& \frac{1}{\xi } \\ 
\frac{-i\gamma }{\xi \xi ^{\ast }} & 0 & \frac{1}{\xi ^{\ast }} & \frac{%
V-\epsilon }{-\xi \xi ^{\ast }}%
\end{array}%
\right) ,
\end{equation}%
where  $\epsilon =\pm \sqrt{V^{2}-\gamma ^{2}}$ and%
\begin{equation}
\partial _{k}\left( h_{k}-\epsilon \right) =\left( 
\begin{array}{cccc}
0 & \frac{\partial \xi }{\partial k} & 0 & 0 \\ 
\frac{\partial \xi ^{\ast }}{\partial k} & 0 & 0 & 0 \\ 
0 & 0 & 0 & \frac{\partial \xi ^{\ast }}{\partial k} \\ 
0 & 0 & \frac{\partial \xi }{\partial k} & 0%
\end{array}%
\right) ,
\end{equation}%
then we have%
\begin{equation}
\mathrm{diag}\left[ \left( h_{k}-\epsilon \right) ^{-1}\partial _{k}\left(
h_{k}-\epsilon \right) \right] =\left( 
\begin{array}{cc}
\Gamma & 0 \\ 
0 & \Gamma ^{\dag }%
\end{array}%
\right) ,
\end{equation}%
where%
\begin{equation}
\Gamma =\left( 
\begin{array}{cc}
\frac{\partial \log \left( \xi ^{\ast }\right) }{\partial k} & 0 \\ 
0 & \frac{\partial \log \left( \xi \right) }{\partial k}%
\end{array}%
\right) .
\end{equation}

According to the definition of the sublattice winding number in Eq. (\ref%
{subwinding}), we have%
\begin{eqnarray}
w_{\epsilon }^{A} &=&\int_{0}^{2\pi }\frac{dk}{2\pi i}\partial _{k}\log
\left( \xi ^{\ast }\right) ,  \notag \\
w_{\epsilon }^{B} &=&\int_{0}^{2\pi }\frac{dk}{2\pi i}\partial _{k}\log
\left( \xi \right).
\end{eqnarray}%
Similarly, we get 
\begin{eqnarray}
w_{\epsilon }^{D} &=&w_{\epsilon }^{A},  \notag \\
w_{\epsilon }^{C} &=&w_{\epsilon }^{B}.
\end{eqnarray}%
Further calculation shows that 
\begin{eqnarray}
w_{\epsilon }^{A} &=&\left\{ 
\begin{array}{c}
0,\delta <0 \\ 
1,\delta >0%
\end{array}%
\right. ,  \notag \\
w_{\epsilon }^{B} &=&\left\{ 
\begin{array}{c}
0,\delta <0 \\ 
-1,\delta >0%
\end{array}%
\right. ,
\end{eqnarray}%
and%
\begin{equation}
w^{A}=2w_{\epsilon }^{A}=\left\{ 
\begin{array}{c}
0,\delta <0 \\ 
2,\delta >0%
\end{array}%
\right. .
\end{equation}

\section*{Appendix B: Solution of the edge states}

\label{B} \setcounter{equation}{0} \renewcommand{\theequation}{B%
\arabic{equation}}

Here we present the detailed derivation of the edge states of the SSH ladder 
$\mathcal{H}$. The ansatz of the edge state can be taken as 
\begin{equation}
\left\vert \psi \right\rangle =\sum_{l=1}^{2N}\left( \alpha _{l}a_{l}^{\dag
}+\beta _{l}b_{l}^{\dag }\right) \left\vert 0\right\rangle .
\end{equation}%
Substituting it and $\mathcal{H}$ into 
\begin{equation}
\mathcal{H}\left\vert \psi \right\rangle =\pm \sqrt{V^{2}-\gamma ^{2}}%
\left\vert \psi \right\rangle ,  \label{Sch}
\end{equation}%
we obtain 
\begin{eqnarray}
\left( 1-\delta \right) \beta _{2j+2}+\left( 1+\delta \right) \beta
_{2j}+\chi _{j+1}^{\pm } &=&0,  \notag \\
\left( 1-\delta \right) \alpha _{2j+2}+\left( 1+\delta \right) \alpha
_{2j}+\kappa _{\pm }\chi _{j+1}^{\pm } &=&0,  \notag \\
\left( 1+\delta \right) \alpha _{2j+1}+\left( 1-\delta \right) \alpha
_{2j-1}+\kappa _{\pm }\eta _{j}^{\pm } &=&0,  \notag \\
\left( 1+\delta \right) \beta _{2j+1}+\left( 1-\delta \right) \beta
_{2j-1}+\eta _{j}^{\pm } &=&0,  \label{bulk}
\end{eqnarray}%
for $j\in \left[ 1,N-1\right] $, and 
\begin{eqnarray}
\left( 1-\delta \right) \beta _{2}+\chi _{1}^{\pm } &=&0,  \notag \\
\left( 1-\delta \right) \alpha _{2}+\kappa _{\pm }\chi _{1}^{\pm } &=&0, 
\notag \\
\left( 1-\delta \right) \alpha _{2N-1}+\kappa _{\pm }\eta _{N}^{\pm } &=&0, 
\notag \\
\left( 1-\delta \right) \beta _{2N-1}+\eta _{N}^{\pm } &=&0,  \label{edge}
\end{eqnarray}%
for the boundary terms. Here 
\begin{eqnarray}
\chi _{j}^{\pm } &=&\left( V-\mathcal{E}_{\pm }\right) \alpha
_{2j-1}+i\gamma \beta _{2j-1},  \notag \\
\eta _{j}^{\pm } &=&\left(\mathcal{E}_{\pm }-V \right) \alpha _{2j}-i\gamma
\beta _{2j},
\end{eqnarray}%
for $j\in \left[ 1,N\right] $, and $\kappa _{\pm }$ are defined as 
\begin{equation}
\kappa _{\pm }=\frac{-i\gamma }{V\mp \sqrt{V^{2}-\gamma ^{2}}}.
\end{equation}

Combining Eq. (\ref{bulk}) and Eq. (\ref{edge}), we can get the following
equations 
\begin{eqnarray}
\left( 1-\delta \right) \beta _{2j+2}+\left( 1+\delta \right) \beta _{2j}
&=&0,  \notag \\
\left( 1+\delta \right) \alpha _{2j+1}+\left( 1-\delta \right) \alpha
_{2j-1} &=&0,
\end{eqnarray}%
for $j\in \left[ 1,N-1\right] $, and 
\begin{equation}
\beta _{2}=0,\alpha _{2N-1}=0.
\end{equation}%
Here 
\begin{equation}
\alpha _{l}=\kappa _{\pm }\beta _{l},
\end{equation}%
with $l\in \left[ 1,2N\right]$.

Thus, under the large $N$ limit and with condition $0<\delta <1$, there are
four edge eigenstates localized at the two ends of the system, which have
the form 
\begin{eqnarray}
\left\vert \psi _{\mathrm{L}}^{\pm }\right\rangle  &=&\sum_{j=1}^{N}\left(
\alpha _{2j-1}a_{2j-1}^{\dag }+\beta _{2j-1}b_{2j-1}^{\dag }\right)
\left\vert 0\right\rangle ,  \notag \\
\left\vert \psi _{\mathrm{R}}^{\pm }\right\rangle  &=&\sum_{j=1}^{N}\left(
\alpha_{2j}a_{2j}^{\dag }+\beta _{2j}b_{2j}^{\dag }\right) \left\vert
0\right\rangle ,
\end{eqnarray}%
where the parameters satisfy 
\begin{eqnarray}
&&\alpha _{2j-1}\left( \beta _{2j-1}\right) =\left( -1\right)
^{j-1}e^{-\left( j-1\right) \zeta }\alpha _{1}\left( \beta _{1}\right) ,j\in 
\left[ 2,N\right] ,  \notag \\
&&\alpha _{2j}\left( \beta _{2j}\right) =\left( -1\right) ^{N-j}e^{-\left(
N-j\right) \zeta }\alpha _{2N}\left( \beta _{2N}\right) ,j\in \left[ 1,N-1%
\right] ,  \notag \\
&&\alpha _{l}=\kappa _{\pm }\beta _{l},l\in \left[ 1,2N\right] ,
\end{eqnarray}%
and the localization length is $\zeta =\ln \left[ \left( \delta +1\right)
/\left( 1-\delta \right) \right] >0\ $ $\left( 0<\delta <1\right) $.

%

\begin{thebibliography}{54}%
\makeatletter
\providecommand \@ifxundefined [1]{%
 \@ifx{#1\undefined}
}%
\providecommand \@ifnum [1]{%
 \ifnum #1\expandafter \@firstoftwo
 \else \expandafter \@secondoftwo
 \fi
}%
\providecommand \@ifx [1]{%
 \ifx #1\expandafter \@firstoftwo
 \else \expandafter \@secondoftwo
 \fi
}%
\providecommand \natexlab [1]{#1}%
\providecommand \enquote  [1]{``#1''}%
\providecommand \bibnamefont  [1]{#1}%
\providecommand \bibfnamefont [1]{#1}%
\providecommand \citenamefont [1]{#1}%
\providecommand \href@noop [0]{\@secondoftwo}%
\providecommand \href [0]{\begingroup \@sanitize@url \@href}%
\providecommand \@href[1]{\@@startlink{#1}\@@href}%
\providecommand \@@href[1]{\endgroup#1\@@endlink}%
\providecommand \@sanitize@url [0]{\catcode `\\12\catcode `\$12\catcode
  `\&12\catcode `\#12\catcode `\^12\catcode `\_12\catcode `\%12\relax}%
\providecommand \@@startlink[1]{}%
\providecommand \@@endlink[0]{}%
\providecommand \url  [0]{\begingroup\@sanitize@url \@url }%
\providecommand \@url [1]{\endgroup\@href {#1}{\urlprefix }}%
\providecommand \urlprefix  [0]{URL }%
\providecommand \Eprint [0]{\href }%
\providecommand \doibase [0]{https://doi.org/}%
\providecommand \selectlanguage [0]{\@gobble}%
\providecommand \bibinfo  [0]{\@secondoftwo}%
\providecommand \bibfield  [0]{\@secondoftwo}%
\providecommand \translation [1]{[#1]}%
\providecommand \BibitemOpen [0]{}%
\providecommand \bibitemStop [0]{}%
\providecommand \bibitemNoStop [0]{.\EOS\space}%
\providecommand \EOS [0]{\spacefactor3000\relax}%
\providecommand \BibitemShut  [1]{\csname bibitem#1\endcsname}%
\let\auto@bib@innerbib\@empty
\bibitem [{\citenamefont {Klitzing}\ \emph {et~al.}(1980)\citenamefont
  {Klitzing}, \citenamefont {Dorda},\ and\ \citenamefont
  {Pepper}}]{klitzing1980}%
  \BibitemOpen
  \bibfield  {author} {\bibinfo {author} {\bibfnamefont {K.~v.}\ \bibnamefont
  {Klitzing}}, \bibinfo {author} {\bibfnamefont {G.}~\bibnamefont {Dorda}},\
  and\ \bibinfo {author} {\bibfnamefont {M.}~\bibnamefont {Pepper}},\
  }\bibfield  {title} {\bibinfo {title} {New method for high-accuracy
  determination of the fine-structure constant based on quantized hall
  resistance},\ }\href {https://doi.org/10.1103/PhysRevLett.45.494} {\bibfield
  {journal} {\bibinfo  {journal} {Phys. Rev. Lett.}\ }\textbf {\bibinfo
  {volume} {45}},\ \bibinfo {pages} {494} (\bibinfo {year} {1980})}\BibitemShut
  {NoStop}%
\bibitem [{\citenamefont {Thouless}\ \emph {et~al.}(1982)\citenamefont
  {Thouless}, \citenamefont {Kohmoto}, \citenamefont {Nightingale},\ and\
  \citenamefont {den Nijs}}]{Thouless1982}%
  \BibitemOpen
  \bibfield  {author} {\bibinfo {author} {\bibfnamefont {D.~J.}\ \bibnamefont
  {Thouless}}, \bibinfo {author} {\bibfnamefont {M.}~\bibnamefont {Kohmoto}},
  \bibinfo {author} {\bibfnamefont {M.~P.}\ \bibnamefont {Nightingale}},\ and\
  \bibinfo {author} {\bibfnamefont {M.}~\bibnamefont {den Nijs}},\ }\bibfield
  {title} {\bibinfo {title} {Quantized hall conductance in a two-dimensional
  periodic potential},\ }\href {https://doi.org/10.1103/PhysRevLett.49.405}
  {\bibfield  {journal} {\bibinfo  {journal} {Phys. Rev. Lett.}\ }\textbf
  {\bibinfo {volume} {49}},\ \bibinfo {pages} {405} (\bibinfo {year}
  {1982})}\BibitemShut {NoStop}%
\bibitem [{\citenamefont {Zak}(1989)}]{Zak1989}%
  \BibitemOpen
  \bibfield  {author} {\bibinfo {author} {\bibfnamefont {J.}~\bibnamefont
  {Zak}},\ }\bibfield  {title} {\bibinfo {title} {Berry’s phase for energy
  bands in solids},\ }\href {https://doi.org/10.1103/PhysRevLett.62.2747}
  {\bibfield  {journal} {\bibinfo  {journal} {Phys. Rev. Lett.}\ }\textbf
  {\bibinfo {volume} {62}},\ \bibinfo {pages} {2747} (\bibinfo {year}
  {1989})}\BibitemShut {NoStop}%
\bibitem [{\citenamefont {Kitaev}(2001)}]{Kitaev2001}%
  \BibitemOpen
  \bibfield  {author} {\bibinfo {author} {\bibfnamefont {A.~Y.}\ \bibnamefont
  {Kitaev}},\ }\bibfield  {title} {\bibinfo {title} {Unpaired majorana fermions
  in quantum wires},\ }\href {https://doi.org/10.1070/1063-7869/44/10s/s29}
  {\bibfield  {journal} {\bibinfo  {journal} {Phys. Usp.}\ }\textbf {\bibinfo
  {volume} {44}},\ \bibinfo {pages} {131} (\bibinfo {year} {2001})}\BibitemShut
  {NoStop}%
\bibitem [{\citenamefont {Fu}\ and\ \citenamefont {Kane}(2007)}]{Fu2007a}%
  \BibitemOpen
  \bibfield  {author} {\bibinfo {author} {\bibfnamefont {L.}~\bibnamefont
  {Fu}}\ and\ \bibinfo {author} {\bibfnamefont {C.~L.}\ \bibnamefont {Kane}},\
  }\bibfield  {title} {\bibinfo {title} {Topological insulators with inversion
  symmetry},\ }\href {https://doi.org/10.1103/PhysRevB.76.045302} {\bibfield
  {journal} {\bibinfo  {journal} {Phys. Rev. B}\ }\textbf {\bibinfo {volume}
  {76}},\ \bibinfo {pages} {045302} (\bibinfo {year} {2007})}\BibitemShut
  {NoStop}%
\bibitem [{\citenamefont {Nayak}\ \emph {et~al.}(2008)\citenamefont {Nayak},
  \citenamefont {Simon}, \citenamefont {Stern}, \citenamefont {Freedman},\ and\
  \citenamefont {Das~Sarma}}]{Nayak2008a}%
  \BibitemOpen
  \bibfield  {author} {\bibinfo {author} {\bibfnamefont {C.}~\bibnamefont
  {Nayak}}, \bibinfo {author} {\bibfnamefont {S.~H.}\ \bibnamefont {Simon}},
  \bibinfo {author} {\bibfnamefont {A.}~\bibnamefont {Stern}}, \bibinfo
  {author} {\bibfnamefont {M.}~\bibnamefont {Freedman}},\ and\ \bibinfo
  {author} {\bibfnamefont {S.}~\bibnamefont {Das~Sarma}},\ }\bibfield  {title}
  {\bibinfo {title} {Non-abelian anyons and topological quantum computation},\
  }\href {https://doi.org/10.1103/RevModPhys.80.1083} {\bibfield  {journal}
  {\bibinfo  {journal} {Rev. Mod. Phys.}\ }\textbf {\bibinfo {volume} {80}},\
  \bibinfo {pages} {1083} (\bibinfo {year} {2008})}\BibitemShut {NoStop}%
\bibitem [{\citenamefont {Qi}\ and\ \citenamefont
  {Zhang}(2011)}]{qi2011topological}%
  \BibitemOpen
  \bibfield  {author} {\bibinfo {author} {\bibfnamefont {X.-L.}\ \bibnamefont
  {Qi}}\ and\ \bibinfo {author} {\bibfnamefont {S.-C.}\ \bibnamefont {Zhang}},\
  }\bibfield  {title} {\bibinfo {title} {Topological insulators and
  superconductors},\ }\href {https://doi.org/10.1103/RevModPhys.83.1057}
  {\bibfield  {journal} {\bibinfo  {journal} {Rev. Mod. Phys.}\ }\textbf
  {\bibinfo {volume} {83}},\ \bibinfo {pages} {1057} (\bibinfo {year}
  {2011})}\BibitemShut {NoStop}%
\bibitem [{\citenamefont {Matsuura}\ \emph {et~al.}(2013)\citenamefont
  {Matsuura}, \citenamefont {Chang}, \citenamefont {Schnyder},\ and\
  \citenamefont {Ryu}}]{matsuura2013protected}%
  \BibitemOpen
  \bibfield  {author} {\bibinfo {author} {\bibfnamefont {S.}~\bibnamefont
  {Matsuura}}, \bibinfo {author} {\bibfnamefont {P.-Y.}\ \bibnamefont {Chang}},
  \bibinfo {author} {\bibfnamefont {A.~P.}\ \bibnamefont {Schnyder}},\ and\
  \bibinfo {author} {\bibfnamefont {S.}~\bibnamefont {Ryu}},\ }\bibfield
  {title} {\bibinfo {title} {Protected boundary states in gapless topological
  phases},\ }\href
  {https://iopscience.iop.org/article/10.1088/1367-2630/15/6/065001} {\bibfield
   {journal} {\bibinfo  {journal} {New J. Phys.}\ }\textbf {\bibinfo {volume}
  {15}},\ \bibinfo {pages} {065001} (\bibinfo {year} {2013})}\BibitemShut
  {NoStop}%
\bibitem [{\citenamefont {Chiu}\ and\ \citenamefont
  {Schnyder}(2014)}]{Chiu2014}%
  \BibitemOpen
  \bibfield  {author} {\bibinfo {author} {\bibfnamefont {C.-K.}\ \bibnamefont
  {Chiu}}\ and\ \bibinfo {author} {\bibfnamefont {A.~P.}\ \bibnamefont
  {Schnyder}},\ }\bibfield  {title} {\bibinfo {title} {Classification of
  reflection-symmetry-protected topological semimetals and nodal
  superconductors},\ }\href {https://doi.org/10.1103/PhysRevB.90.205136}
  {\bibfield  {journal} {\bibinfo  {journal} {Phys. Rev. B}\ }\textbf {\bibinfo
  {volume} {90}},\ \bibinfo {pages} {205136} (\bibinfo {year}
  {2014})}\BibitemShut {NoStop}%
\bibitem [{\citenamefont {Sarma}\ \emph {et~al.}(2015)\citenamefont {Sarma},
  \citenamefont {Freedman},\ and\ \citenamefont {Nayak}}]{Sarma2015}%
  \BibitemOpen
  \bibfield  {author} {\bibinfo {author} {\bibfnamefont {S.~D.}\ \bibnamefont
  {Sarma}}, \bibinfo {author} {\bibfnamefont {M.}~\bibnamefont {Freedman}},\
  and\ \bibinfo {author} {\bibfnamefont {C.}~\bibnamefont {Nayak}},\ }\bibfield
   {title} {\bibinfo {title} {Majorana zero modes and topological quantum
  computation},\ }\href {https://doi.org/10.1038/npjqi.2015.1} {\bibfield
  {journal} {\bibinfo  {journal} {npj Quantum Information}\ }\textbf {\bibinfo
  {volume} {1}},\ \bibinfo {pages} {1} (\bibinfo {year} {2015})}\BibitemShut
  {NoStop}%
\bibitem [{\citenamefont {Chiu}\ \emph {et~al.}(2016)\citenamefont {Chiu},
  \citenamefont {Teo}, \citenamefont {Schnyder},\ and\ \citenamefont
  {Ryu}}]{Chiu2016}%
  \BibitemOpen
  \bibfield  {author} {\bibinfo {author} {\bibfnamefont {C.-K.}\ \bibnamefont
  {Chiu}}, \bibinfo {author} {\bibfnamefont {J.~C.~Y.}\ \bibnamefont {Teo}},
  \bibinfo {author} {\bibfnamefont {A.~P.}\ \bibnamefont {Schnyder}},\ and\
  \bibinfo {author} {\bibfnamefont {S.}~\bibnamefont {Ryu}},\ }\bibfield
  {title} {\bibinfo {title} {Classification of topological quantum matter with
  symmetries},\ }\href {https://doi.org/10.1103/RevModPhys.88.035005}
  {\bibfield  {journal} {\bibinfo  {journal} {Rev. Mod. Phys.}\ }\textbf
  {\bibinfo {volume} {88}},\ \bibinfo {pages} {035005} (\bibinfo {year}
  {2016})}\BibitemShut {NoStop}%
\bibitem [{\citenamefont {Asb{\'o}th}\ \emph {et~al.}(2016)\citenamefont
  {Asb{\'o}th}, \citenamefont {Oroszl{\'a}ny},\ and\ \citenamefont
  {P{\'a}lyi}}]{Asboth2016}%
  \BibitemOpen
  \bibfield  {author} {\bibinfo {author} {\bibfnamefont {J.~K.}\ \bibnamefont
  {Asb{\'o}th}}, \bibinfo {author} {\bibfnamefont {L.}~\bibnamefont
  {Oroszl{\'a}ny}},\ and\ \bibinfo {author} {\bibfnamefont {A.}~\bibnamefont
  {P{\'a}lyi}},\ }\href@noop {} {\emph {\bibinfo {title} {A short course on
  topological insulators}}}\ (\bibinfo  {publisher} {Springer},\ \bibinfo
  {year} {2016})\BibitemShut {NoStop}%
\bibitem [{\citenamefont {Kobayashi}\ \emph {et~al.}(2018)\citenamefont
  {Kobayashi}, \citenamefont {Sumita}, \citenamefont {Yanase},\ and\
  \citenamefont {Sato}}]{kobayashi2018symmetry}%
  \BibitemOpen
  \bibfield  {author} {\bibinfo {author} {\bibfnamefont {S.}~\bibnamefont
  {Kobayashi}}, \bibinfo {author} {\bibfnamefont {S.}~\bibnamefont {Sumita}},
  \bibinfo {author} {\bibfnamefont {Y.}~\bibnamefont {Yanase}},\ and\ \bibinfo
  {author} {\bibfnamefont {M.}~\bibnamefont {Sato}},\ }\bibfield  {title}
  {\bibinfo {title} {Symmetry-protected line nodes and majorana flat bands in
  nodal crystalline superconductors},\ }\href
  {https://doi.org/10.1103/PhysRevB.97.180504} {\bibfield  {journal} {\bibinfo
  {journal} {Phys. Rev. B}\ }\textbf {\bibinfo {volume} {97}},\ \bibinfo
  {pages} {180504} (\bibinfo {year} {2018})}\BibitemShut {NoStop}%
\bibitem [{\citenamefont {Xie}\ \emph {et~al.}(2023)\citenamefont {Xie},
  \citenamefont {Jin},\ and\ \citenamefont {Song}}]{xie2023antihelical}%
  \BibitemOpen
  \bibfield  {author} {\bibinfo {author} {\bibfnamefont {L.}~\bibnamefont
  {Xie}}, \bibinfo {author} {\bibfnamefont {L.}~\bibnamefont {Jin}},\ and\
  \bibinfo {author} {\bibfnamefont {Z.}~\bibnamefont {Song}},\ }\bibfield
  {title} {\bibinfo {title} {Antihelical edge states in two-dimensional
  photonic topological metals},\ }\href
  {https://doi.org/https://doi.org/10.1016/j.scib.2023.01.018} {\bibfield
  {journal} {\bibinfo  {journal} {Sci. Bull.}\ }\textbf {\bibinfo {volume}
  {68}},\ \bibinfo {pages} {255} (\bibinfo {year} {2023})}\BibitemShut
  {NoStop}%
\bibitem [{\citenamefont {Rhim}\ \emph {et~al.}(2017)\citenamefont {Rhim},
  \citenamefont {Behrends},\ and\ \citenamefont {Bardarson}}]{rhim2017bulk}%
  \BibitemOpen
  \bibfield  {author} {\bibinfo {author} {\bibfnamefont {J.-W.}\ \bibnamefont
  {Rhim}}, \bibinfo {author} {\bibfnamefont {J.}~\bibnamefont {Behrends}},\
  and\ \bibinfo {author} {\bibfnamefont {J.~H.}\ \bibnamefont {Bardarson}},\
  }\bibfield  {title} {\bibinfo {title} {Bulk-boundary correspondence from the
  intercellular zak phase},\ }\href
  {https://doi.org/10.1103/PhysRevB.95.035421} {\bibfield  {journal} {\bibinfo
  {journal} {Phys. Rev. B}\ }\textbf {\bibinfo {volume} {95}},\ \bibinfo
  {pages} {035421} (\bibinfo {year} {2017})}\BibitemShut {NoStop}%
\bibitem [{\citenamefont {Guzmán}\ \emph {et~al.}(2022)\citenamefont
  {Guzmán}, \citenamefont {Bartolo},\ and\ \citenamefont
  {Carpentier}}]{guzman2022geometry}%
  \BibitemOpen
  \bibfield  {author} {\bibinfo {author} {\bibfnamefont {M.}~\bibnamefont
  {Guzmán}}, \bibinfo {author} {\bibfnamefont {D.}~\bibnamefont {Bartolo}},\
  and\ \bibinfo {author} {\bibfnamefont {D.}~\bibnamefont {Carpentier}},\
  }\bibfield  {title} {\bibinfo {title} {{Geometry and topology tango in
  ordered and amorphous chiral matter}},\ }\href
  {https://doi.org/10.21468/SciPostPhys.12.1.038} {\bibfield  {journal}
  {\bibinfo  {journal} {SciPost Phys.}\ }\textbf {\bibinfo {volume} {12}},\
  \bibinfo {pages} {038} (\bibinfo {year} {2022})}\BibitemShut {NoStop}%
\bibitem [{\citenamefont {Verma}\ and\ \citenamefont
  {Ghosh}(2024)}]{verma2024bulk}%
  \BibitemOpen
  \bibfield  {author} {\bibinfo {author} {\bibfnamefont {S.}~\bibnamefont
  {Verma}}\ and\ \bibinfo {author} {\bibfnamefont {T.~K.}\ \bibnamefont
  {Ghosh}},\ }\bibfield  {title} {\bibinfo {title} {Bulk-boundary
  correspondence in extended trimer su-schrieffer-heeger model},\ }\href
  {https://doi.org/10.1103/PhysRevB.110.125424} {\bibfield  {journal} {\bibinfo
   {journal} {Phys. Rev. B}\ }\textbf {\bibinfo {volume} {110}},\ \bibinfo
  {pages} {125424} (\bibinfo {year} {2024})}\BibitemShut {NoStop}%
\bibitem [{\citenamefont {Else}\ \emph {et~al.}(2017)\citenamefont {Else},
  \citenamefont {Fendley}, \citenamefont {Kemp},\ and\ \citenamefont
  {Nayak}}]{Else2017}%
  \BibitemOpen
  \bibfield  {author} {\bibinfo {author} {\bibfnamefont {D.~V.}\ \bibnamefont
  {Else}}, \bibinfo {author} {\bibfnamefont {P.}~\bibnamefont {Fendley}},
  \bibinfo {author} {\bibfnamefont {J.}~\bibnamefont {Kemp}},\ and\ \bibinfo
  {author} {\bibfnamefont {C.}~\bibnamefont {Nayak}},\ }\bibfield  {title}
  {\bibinfo {title} {Prethermal strong zero modes and topological qubits},\
  }\href {https://doi.org/10.1103/PhysRevX.7.041062} {\bibfield  {journal}
  {\bibinfo  {journal} {Phys. Rev. X}\ }\textbf {\bibinfo {volume} {7}},\
  \bibinfo {pages} {041062} (\bibinfo {year} {2017})}\BibitemShut {NoStop}%
\bibitem [{\citenamefont {Rechtsman}\ \emph {et~al.}(2013)\citenamefont
  {Rechtsman}, \citenamefont {Zeuner}, \citenamefont {Plotnik}, \citenamefont
  {Lumer}, \citenamefont {Podolsky}, \citenamefont {Dreisow}, \citenamefont
  {Nolte}, \citenamefont {Segev},\ and\ \citenamefont
  {Szameit}}]{rechtsman2013photonic}%
  \BibitemOpen
  \bibfield  {author} {\bibinfo {author} {\bibfnamefont {M.~C.}\ \bibnamefont
  {Rechtsman}}, \bibinfo {author} {\bibfnamefont {J.~M.}\ \bibnamefont
  {Zeuner}}, \bibinfo {author} {\bibfnamefont {Y.}~\bibnamefont {Plotnik}},
  \bibinfo {author} {\bibfnamefont {Y.}~\bibnamefont {Lumer}}, \bibinfo
  {author} {\bibfnamefont {D.}~\bibnamefont {Podolsky}}, \bibinfo {author}
  {\bibfnamefont {F.}~\bibnamefont {Dreisow}}, \bibinfo {author} {\bibfnamefont
  {S.}~\bibnamefont {Nolte}}, \bibinfo {author} {\bibfnamefont
  {M.}~\bibnamefont {Segev}},\ and\ \bibinfo {author} {\bibfnamefont
  {A.}~\bibnamefont {Szameit}},\ }\bibfield  {title} {\bibinfo {title}
  {Photonic floquet topological insulators},\ }\href@noop {} {\bibfield
  {journal} {\bibinfo  {journal} {Nature}\ }\textbf {\bibinfo {volume} {496}},\
  \bibinfo {pages} {196} (\bibinfo {year} {2013})}\BibitemShut {NoStop}%
\bibitem [{\citenamefont {Lohse}\ \emph {et~al.}(2018)\citenamefont {Lohse},
  \citenamefont {Schweizer}, \citenamefont {Price}, \citenamefont
  {Zilberberg},\ and\ \citenamefont {Bloch}}]{Lohse2018}%
  \BibitemOpen
  \bibfield  {author} {\bibinfo {author} {\bibfnamefont {M.}~\bibnamefont
  {Lohse}}, \bibinfo {author} {\bibfnamefont {C.}~\bibnamefont {Schweizer}},
  \bibinfo {author} {\bibfnamefont {H.~M.}\ \bibnamefont {Price}}, \bibinfo
  {author} {\bibfnamefont {O.}~\bibnamefont {Zilberberg}},\ and\ \bibinfo
  {author} {\bibfnamefont {I.}~\bibnamefont {Bloch}},\ }\bibfield  {title}
  {\bibinfo {title} {Exploring 4d quantum hall physics with a 2d topological
  charge pump},\ }\href {https://www.nature.com/articles/nature25000}
  {\bibfield  {journal} {\bibinfo  {journal} {Nature}\ }\textbf {\bibinfo
  {volume} {553}},\ \bibinfo {pages} {55} (\bibinfo {year} {2018})}\BibitemShut
  {NoStop}%
\bibitem [{\citenamefont {Sun}\ \emph {et~al.}(2018)\citenamefont {Sun},
  \citenamefont {Yi}, \citenamefont {Wang}, \citenamefont {Zhang},
  \citenamefont {Sanders}, \citenamefont {Xu}, \citenamefont {Wang},
  \citenamefont {Schmiedmayer}, \citenamefont {Deng}, \citenamefont {Liu},
  \citenamefont {Chen},\ and\ \citenamefont {Pan}}]{sun2018}%
  \BibitemOpen
  \bibfield  {author} {\bibinfo {author} {\bibfnamefont {W.}~\bibnamefont
  {Sun}}, \bibinfo {author} {\bibfnamefont {C.-R.}\ \bibnamefont {Yi}},
  \bibinfo {author} {\bibfnamefont {B.-Z.}\ \bibnamefont {Wang}}, \bibinfo
  {author} {\bibfnamefont {W.-W.}\ \bibnamefont {Zhang}}, \bibinfo {author}
  {\bibfnamefont {B.~C.}\ \bibnamefont {Sanders}}, \bibinfo {author}
  {\bibfnamefont {X.-T.}\ \bibnamefont {Xu}}, \bibinfo {author} {\bibfnamefont
  {Z.-Y.}\ \bibnamefont {Wang}}, \bibinfo {author} {\bibfnamefont
  {J.}~\bibnamefont {Schmiedmayer}}, \bibinfo {author} {\bibfnamefont
  {Y.}~\bibnamefont {Deng}}, \bibinfo {author} {\bibfnamefont {X.-J.}\
  \bibnamefont {Liu}}, \bibinfo {author} {\bibfnamefont {S.}~\bibnamefont
  {Chen}},\ and\ \bibinfo {author} {\bibfnamefont {J.-W.}\ \bibnamefont
  {Pan}},\ }\bibfield  {title} {\bibinfo {title} {Uncover topology by quantum
  quench dynamics},\ }\href {https://doi.org/10.1103/PhysRevLett.121.250403}
  {\bibfield  {journal} {\bibinfo  {journal} {Phys. Rev. Lett.}\ }\textbf
  {\bibinfo {volume} {121}},\ \bibinfo {pages} {250403} (\bibinfo {year}
  {2018})}\BibitemShut {NoStop}%
\bibitem [{\citenamefont {Zhang}\ \emph {et~al.}(2018)\citenamefont {Zhang},
  \citenamefont {Zhu}, \citenamefont {Zhao}, \citenamefont {Yan},\ and\
  \citenamefont {Zhu}}]{zhang2018a}%
  \BibitemOpen
  \bibfield  {author} {\bibinfo {author} {\bibfnamefont {D.-W.}\ \bibnamefont
  {Zhang}}, \bibinfo {author} {\bibfnamefont {Y.-Q.}\ \bibnamefont {Zhu}},
  \bibinfo {author} {\bibfnamefont {Y.}~\bibnamefont {Zhao}}, \bibinfo {author}
  {\bibfnamefont {H.}~\bibnamefont {Yan}},\ and\ \bibinfo {author}
  {\bibfnamefont {S.-L.}\ \bibnamefont {Zhu}},\ }\bibfield  {title} {\bibinfo
  {title} {Topological quantum matter with cold atoms},\ }\href
  {https://www.tandfonline.com/doi/abs/10.1080/00018732.2019.1594094}
  {\bibfield  {journal} {\bibinfo  {journal} {Adv. Phys.}\ }\textbf {\bibinfo
  {volume} {67}},\ \bibinfo {pages} {253} (\bibinfo {year} {2018})}\BibitemShut
  {NoStop}%
\bibitem [{\citenamefont {Li}\ \emph {et~al.}(2019)\citenamefont {Li},
  \citenamefont {Harter}, \citenamefont {Liu}, \citenamefont {de~Melo},
  \citenamefont {Joglekar},\ and\ \citenamefont {Luo}}]{Li2019}%
  \BibitemOpen
  \bibfield  {author} {\bibinfo {author} {\bibfnamefont {J.}~\bibnamefont
  {Li}}, \bibinfo {author} {\bibfnamefont {A.~K.}\ \bibnamefont {Harter}},
  \bibinfo {author} {\bibfnamefont {J.}~\bibnamefont {Liu}}, \bibinfo {author}
  {\bibfnamefont {L.}~\bibnamefont {de~Melo}}, \bibinfo {author} {\bibfnamefont
  {Y.~N.}\ \bibnamefont {Joglekar}},\ and\ \bibinfo {author} {\bibfnamefont
  {L.}~\bibnamefont {Luo}},\ }\bibfield  {title} {\bibinfo {title} {Observation
  of parity-time symmetry breaking transitions in a dissipative {F}loquet
  system of ultracold atoms},\ }\href
  {https://www.nature.com/articles/s41467-019-08596-1} {\bibfield  {journal}
  {\bibinfo  {journal} {Nat. Commun.}\ }\textbf {\bibinfo {volume} {10}},\
  \bibinfo {pages} {855} (\bibinfo {year} {2019})}\BibitemShut {NoStop}%
\bibitem [{\citenamefont {Barbarino}\ \emph {et~al.}(2020)\citenamefont
  {Barbarino}, \citenamefont {Yu}, \citenamefont {Zoller},\ and\ \citenamefont
  {Budich}}]{barbarino2020preparing}%
  \BibitemOpen
  \bibfield  {author} {\bibinfo {author} {\bibfnamefont {S.}~\bibnamefont
  {Barbarino}}, \bibinfo {author} {\bibfnamefont {J.}~\bibnamefont {Yu}},
  \bibinfo {author} {\bibfnamefont {P.}~\bibnamefont {Zoller}},\ and\ \bibinfo
  {author} {\bibfnamefont {J.~C.}\ \bibnamefont {Budich}},\ }\bibfield  {title}
  {\bibinfo {title} {Preparing atomic topological quantum matter by adiabatic
  nonunitary dynamics},\ }\href
  {https://doi.org/10.1103/PhysRevLett.124.010401} {\bibfield  {journal}
  {\bibinfo  {journal} {Phys. Rev. Lett.}\ }\textbf {\bibinfo {volume} {124}},\
  \bibinfo {pages} {010401} (\bibinfo {year} {2020})}\BibitemShut {NoStop}%
\bibitem [{\citenamefont {Rudner}\ and\ \citenamefont
  {Lindner}(2020)}]{rudner2020}%
  \BibitemOpen
  \bibfield  {author} {\bibinfo {author} {\bibfnamefont {M.~S.}\ \bibnamefont
  {Rudner}}\ and\ \bibinfo {author} {\bibfnamefont {N.~H.}\ \bibnamefont
  {Lindner}},\ }\bibfield  {title} {\bibinfo {title} {Band structure
  engineering and non-equilibrium dynamics in floquet topological insulators},\
  }\href {https://www.nature.com/articles/s42254-020-0170-z} {\bibfield
  {journal} {\bibinfo  {journal} {Nat. Rev. Phys.}\ }\textbf {\bibinfo {volume}
  {2}},\ \bibinfo {pages} {229} (\bibinfo {year} {2020})}\BibitemShut {NoStop}%
\bibitem [{\citenamefont {Zhang}\ \emph {et~al.}(2020)\citenamefont {Zhang},
  \citenamefont {Zhang},\ and\ \citenamefont {Liu}}]{zhang2020unified}%
  \BibitemOpen
  \bibfield  {author} {\bibinfo {author} {\bibfnamefont {L.}~\bibnamefont
  {Zhang}}, \bibinfo {author} {\bibfnamefont {L.}~\bibnamefont {Zhang}},\ and\
  \bibinfo {author} {\bibfnamefont {X.-J.}\ \bibnamefont {Liu}},\ }\bibfield
  {title} {\bibinfo {title} {Unified theory to characterize floquet topological
  phases by quench dynamics},\ }\href
  {https://doi.org/10.1103/PhysRevLett.125.183001} {\bibfield  {journal}
  {\bibinfo  {journal} {Phys. Rev. Lett.}\ }\textbf {\bibinfo {volume} {125}},\
  \bibinfo {pages} {183001} (\bibinfo {year} {2020})}\BibitemShut {NoStop}%
\bibitem [{\citenamefont {Citro}\ and\ \citenamefont
  {Aidelsburger}(2023)}]{citro2023thouless}%
  \BibitemOpen
  \bibfield  {author} {\bibinfo {author} {\bibfnamefont {R.}~\bibnamefont
  {Citro}}\ and\ \bibinfo {author} {\bibfnamefont {M.}~\bibnamefont
  {Aidelsburger}},\ }\bibfield  {title} {\bibinfo {title} {Thouless pumping and
  topology},\ }\href@noop {} {\bibfield  {journal} {\bibinfo  {journal} {Nat.
  Rev. Phys.}\ }\textbf {\bibinfo {volume} {5}},\ \bibinfo {pages} {87}
  (\bibinfo {year} {2023})}\BibitemShut {NoStop}%
\bibitem [{\citenamefont {Bin}\ \emph {et~al.}(2023)\citenamefont {Bin},
  \citenamefont {Wan}, \citenamefont {Nori}, \citenamefont {Wu},\ and\
  \citenamefont {L\"u}}]{bin2023out}%
  \BibitemOpen
  \bibfield  {author} {\bibinfo {author} {\bibfnamefont {Q.}~\bibnamefont
  {Bin}}, \bibinfo {author} {\bibfnamefont {L.-L.}\ \bibnamefont {Wan}},
  \bibinfo {author} {\bibfnamefont {F.}~\bibnamefont {Nori}}, \bibinfo {author}
  {\bibfnamefont {Y.}~\bibnamefont {Wu}},\ and\ \bibinfo {author}
  {\bibfnamefont {X.-Y.}\ \bibnamefont {L\"u}},\ }\bibfield  {title} {\bibinfo
  {title} {Out-of-time-order correlation as a witness for topological phase
  transitions},\ }\href {https://doi.org/10.1103/PhysRevB.107.L020202}
  {\bibfield  {journal} {\bibinfo  {journal} {Phys. Rev. B}\ }\textbf {\bibinfo
  {volume} {107}},\ \bibinfo {pages} {L020202} (\bibinfo {year}
  {2023})}\BibitemShut {NoStop}%
\bibitem [{\citenamefont {Lane}\ \emph {et~al.}(2024)\citenamefont {Lane},
  \citenamefont {Horv\'ath},\ and\ \citenamefont {Patrick}}]{lane2024extended}%
  \BibitemOpen
  \bibfield  {author} {\bibinfo {author} {\bibfnamefont {T.~L.~M.}\
  \bibnamefont {Lane}}, \bibinfo {author} {\bibfnamefont {M.}~\bibnamefont
  {Horv\'ath}},\ and\ \bibinfo {author} {\bibfnamefont {K.}~\bibnamefont
  {Patrick}},\ }\bibfield  {title} {\bibinfo {title} {Extended edge modes and
  disorder preservation of a symmetry-protected topological phase out of
  equilibrium},\ }\href {https://doi.org/10.1103/PhysRevB.110.165139}
  {\bibfield  {journal} {\bibinfo  {journal} {Phys. Rev. B}\ }\textbf {\bibinfo
  {volume} {110}},\ \bibinfo {pages} {165139} (\bibinfo {year}
  {2024})}\BibitemShut {NoStop}%
\bibitem [{\citenamefont {Jin}(2017)}]{Jin2017}%
  \BibitemOpen
  \bibfield  {author} {\bibinfo {author} {\bibfnamefont {L.}~\bibnamefont
  {Jin}},\ }\bibfield  {title} {\bibinfo {title} {Topological phases and edge
  states in a non-hermitian trimerized optical lattice},\ }\href
  {https://doi.org/10.1103/PhysRevA.96.032103} {\bibfield  {journal} {\bibinfo
  {journal} {Phys. Rev. A}\ }\textbf {\bibinfo {volume} {96}},\ \bibinfo
  {pages} {032103} (\bibinfo {year} {2017})}\BibitemShut {NoStop}%
\bibitem [{\citenamefont {Takata}\ and\ \citenamefont
  {Notomi}(2018)}]{Takata2018}%
  \BibitemOpen
  \bibfield  {author} {\bibinfo {author} {\bibfnamefont {K.}~\bibnamefont
  {Takata}}\ and\ \bibinfo {author} {\bibfnamefont {M.}~\bibnamefont
  {Notomi}},\ }\bibfield  {title} {\bibinfo {title} {Photonic topological
  insulating phase induced solely by gain and loss},\ }\href
  {https://doi.org/10.1103/PhysRevLett.121.213902} {\bibfield  {journal}
  {\bibinfo  {journal} {Phys. Rev. Lett.}\ }\textbf {\bibinfo {volume} {121}},\
  \bibinfo {pages} {213902} (\bibinfo {year} {2018})}\BibitemShut {NoStop}%
\bibitem [{\citenamefont {Harari}\ \emph {et~al.}(2018)\citenamefont {Harari},
  \citenamefont {Bandres}, \citenamefont {Lumer}, \citenamefont {Rechtsman},
  \citenamefont {Chong}, \citenamefont {Khajavikhan}, \citenamefont
  {Christodoulides},\ and\ \citenamefont {Segev}}]{Harari2018}%
  \BibitemOpen
  \bibfield  {author} {\bibinfo {author} {\bibfnamefont {G.}~\bibnamefont
  {Harari}}, \bibinfo {author} {\bibfnamefont {M.~A.}\ \bibnamefont {Bandres}},
  \bibinfo {author} {\bibfnamefont {Y.}~\bibnamefont {Lumer}}, \bibinfo
  {author} {\bibfnamefont {M.~C.}\ \bibnamefont {Rechtsman}}, \bibinfo {author}
  {\bibfnamefont {Y.~D.}\ \bibnamefont {Chong}}, \bibinfo {author}
  {\bibfnamefont {M.}~\bibnamefont {Khajavikhan}}, \bibinfo {author}
  {\bibfnamefont {D.~N.}\ \bibnamefont {Christodoulides}},\ and\ \bibinfo
  {author} {\bibfnamefont {M.}~\bibnamefont {Segev}},\ }\bibfield  {title}
  {\bibinfo {title} {Topological insulator laser: theory},\ }\bibfield
  {journal} {\bibinfo  {journal} {Science}\ }\textbf {\bibinfo {volume}
  {359}},\ \href {https://doi.org/10.1126/science.aar4003}
  {10.1126/science.aar4003} (\bibinfo {year} {2018})\BibitemShut {NoStop}%
\bibitem [{\citenamefont {Bandres}\ \emph {et~al.}(2018)\citenamefont
  {Bandres}, \citenamefont {Wittek}, \citenamefont {Harari}, \citenamefont
  {Parto}, \citenamefont {Ren}, \citenamefont {Segev}, \citenamefont
  {Christodoulides},\ and\ \citenamefont {Khajavikhan}}]{Bandres2018}%
  \BibitemOpen
  \bibfield  {author} {\bibinfo {author} {\bibfnamefont {M.~A.}\ \bibnamefont
  {Bandres}}, \bibinfo {author} {\bibfnamefont {S.}~\bibnamefont {Wittek}},
  \bibinfo {author} {\bibfnamefont {G.}~\bibnamefont {Harari}}, \bibinfo
  {author} {\bibfnamefont {M.}~\bibnamefont {Parto}}, \bibinfo {author}
  {\bibfnamefont {J.}~\bibnamefont {Ren}}, \bibinfo {author} {\bibfnamefont
  {M.}~\bibnamefont {Segev}}, \bibinfo {author} {\bibfnamefont {D.~N.}\
  \bibnamefont {Christodoulides}},\ and\ \bibinfo {author} {\bibfnamefont
  {M.}~\bibnamefont {Khajavikhan}},\ }\bibfield  {title} {\bibinfo {title}
  {Topological insulator laser: Experiments},\ }\bibfield  {journal} {\bibinfo
  {journal} {Science}\ }\textbf {\bibinfo {volume} {359}},\ \href
  {https://doi.org/10.1126/science.aar4005} {10.1126/science.aar4005} (\bibinfo
  {year} {2018})\BibitemShut {NoStop}%
\bibitem [{\citenamefont {Yao}\ \emph {et~al.}(2018)\citenamefont {Yao},
  \citenamefont {Song},\ and\ \citenamefont {Wang}}]{Yao2018a}%
  \BibitemOpen
  \bibfield  {author} {\bibinfo {author} {\bibfnamefont {S.}~\bibnamefont
  {Yao}}, \bibinfo {author} {\bibfnamefont {F.}~\bibnamefont {Song}},\ and\
  \bibinfo {author} {\bibfnamefont {Z.}~\bibnamefont {Wang}},\ }\bibfield
  {title} {\bibinfo {title} {Non-hermitian chern bands},\ }\href
  {https://doi.org/10.1103/PhysRevLett.121.136802} {\bibfield  {journal}
  {\bibinfo  {journal} {Phys. Rev. Lett.}\ }\textbf {\bibinfo {volume} {121}},\
  \bibinfo {pages} {136802} (\bibinfo {year} {2018})}\BibitemShut {NoStop}%
\bibitem [{\citenamefont {Yao}\ and\ \citenamefont {Wang}(2018)}]{Yao2018}%
  \BibitemOpen
  \bibfield  {author} {\bibinfo {author} {\bibfnamefont {S.}~\bibnamefont
  {Yao}}\ and\ \bibinfo {author} {\bibfnamefont {Z.}~\bibnamefont {Wang}},\
  }\bibfield  {title} {\bibinfo {title} {Edge states and topological invariants
  of non-hermitian systems},\ }\href
  {https://doi.org/10.1103/PhysRevLett.121.086803} {\bibfield  {journal}
  {\bibinfo  {journal} {Phys. Rev. Lett.}\ }\textbf {\bibinfo {volume} {121}},\
  \bibinfo {pages} {086803} (\bibinfo {year} {2018})}\BibitemShut {NoStop}%
\bibitem [{\citenamefont {Jin}\ and\ \citenamefont {Song}(2019)}]{Jin2019}%
  \BibitemOpen
  \bibfield  {author} {\bibinfo {author} {\bibfnamefont {L.}~\bibnamefont
  {Jin}}\ and\ \bibinfo {author} {\bibfnamefont {Z.}~\bibnamefont {Song}},\
  }\bibfield  {title} {\bibinfo {title} {Bulk-boundary correspondence in a
  non-hermitian system in one dimension with chiral inversion symmetry},\
  }\href {https://doi.org/10.1103/PhysRevB.99.081103} {\bibfield  {journal}
  {\bibinfo  {journal} {Phys. Rev. B}\ }\textbf {\bibinfo {volume} {99}},\
  \bibinfo {pages} {081103} (\bibinfo {year} {2019})}\BibitemShut {NoStop}%
\bibitem [{\citenamefont {Lee}\ and\ \citenamefont {Thomale}(2019)}]{Lee2019a}%
  \BibitemOpen
  \bibfield  {author} {\bibinfo {author} {\bibfnamefont {C.~H.}\ \bibnamefont
  {Lee}}\ and\ \bibinfo {author} {\bibfnamefont {R.}~\bibnamefont {Thomale}},\
  }\bibfield  {title} {\bibinfo {title} {Anatomy of skin modes and topology in
  non-hermitian systems},\ }\href {https://doi.org/10.1103/PhysRevB.99.201103}
  {\bibfield  {journal} {\bibinfo  {journal} {Phys. Rev. B}\ }\textbf {\bibinfo
  {volume} {99}},\ \bibinfo {pages} {201103} (\bibinfo {year}
  {2019})}\BibitemShut {NoStop}%
\bibitem [{\citenamefont {Luo}\ and\ \citenamefont {Zhang}(2019)}]{Luo2019}%
  \BibitemOpen
  \bibfield  {author} {\bibinfo {author} {\bibfnamefont {X.-W.}\ \bibnamefont
  {Luo}}\ and\ \bibinfo {author} {\bibfnamefont {C.}~\bibnamefont {Zhang}},\
  }\bibfield  {title} {\bibinfo {title} {Higher-order topological corner states
  induced by gain and loss},\ }\href
  {https://doi.org/10.1103/PhysRevLett.123.073601} {\bibfield  {journal}
  {\bibinfo  {journal} {Phys. Rev. Lett.}\ }\textbf {\bibinfo {volume} {123}},\
  \bibinfo {pages} {073601} (\bibinfo {year} {2019})}\BibitemShut {NoStop}%
\bibitem [{\citenamefont {Zhang}\ and\ \citenamefont
  {Gong}(2020)}]{zhang2020non}%
  \BibitemOpen
  \bibfield  {author} {\bibinfo {author} {\bibfnamefont {X.}~\bibnamefont
  {Zhang}}\ and\ \bibinfo {author} {\bibfnamefont {J.}~\bibnamefont {Gong}},\
  }\bibfield  {title} {\bibinfo {title} {Non-hermitian floquet topological
  phases: Exceptional points, coalescent edge modes, and the skin effect},\
  }\href {https://doi.org/10.1103/PhysRevB.101.045415} {\bibfield  {journal}
  {\bibinfo  {journal} {Phys. Rev. B}\ }\textbf {\bibinfo {volume} {101}},\
  \bibinfo {pages} {045415} (\bibinfo {year} {2020})}\BibitemShut {NoStop}%
\bibitem [{\citenamefont {Zhou}\ and\ \citenamefont {Du}(2021)}]{zhou2021non}%
  \BibitemOpen
  \bibfield  {author} {\bibinfo {author} {\bibfnamefont {L.}~\bibnamefont
  {Zhou}}\ and\ \bibinfo {author} {\bibfnamefont {Q.}~\bibnamefont {Du}},\
  }\bibfield  {title} {\bibinfo {title} {Non-hermitian topological phases and
  dynamical quantum phase transitions: A generic connection},\ }\href@noop {}
  {\bibfield  {journal} {\bibinfo  {journal} {New J. Phys.}\ }\textbf {\bibinfo
  {volume} {23}},\ \bibinfo {pages} {063041} (\bibinfo {year}
  {2021})}\BibitemShut {NoStop}%
\bibitem [{\citenamefont {Wu}\ \emph {et~al.}(2021)\citenamefont {Wu},
  \citenamefont {Jin},\ and\ \citenamefont {Song}}]{Wu2021}%
  \BibitemOpen
  \bibfield  {author} {\bibinfo {author} {\bibfnamefont {H.~C.}\ \bibnamefont
  {Wu}}, \bibinfo {author} {\bibfnamefont {L.}~\bibnamefont {Jin}},\ and\
  \bibinfo {author} {\bibfnamefont {Z.}~\bibnamefont {Song}},\ }\bibfield
  {title} {\bibinfo {title} {Topology of an anti-parity-time symmetric
  non-hermitian su-schrieffer-heeger model},\ }\href
  {https://doi.org/10.1103/PhysRevB.103.235110} {\bibfield  {journal} {\bibinfo
   {journal} {Phys. Rev. B}\ }\textbf {\bibinfo {volume} {103}},\ \bibinfo
  {pages} {235110} (\bibinfo {year} {2021})}\BibitemShut {NoStop}%
\bibitem [{\citenamefont {Roccati}\ \emph {et~al.}(2024)\citenamefont
  {Roccati}, \citenamefont {Bello}, \citenamefont {Gong}, \citenamefont {Ueda},
  \citenamefont {Ciccarello}, \citenamefont {Chenu},\ and\ \citenamefont
  {Carollo}}]{roccati2024hermitian}%
  \BibitemOpen
  \bibfield  {author} {\bibinfo {author} {\bibfnamefont {F.}~\bibnamefont
  {Roccati}}, \bibinfo {author} {\bibfnamefont {M.}~\bibnamefont {Bello}},
  \bibinfo {author} {\bibfnamefont {Z.}~\bibnamefont {Gong}}, \bibinfo {author}
  {\bibfnamefont {M.}~\bibnamefont {Ueda}}, \bibinfo {author} {\bibfnamefont
  {F.}~\bibnamefont {Ciccarello}}, \bibinfo {author} {\bibfnamefont
  {A.}~\bibnamefont {Chenu}},\ and\ \bibinfo {author} {\bibfnamefont
  {A.}~\bibnamefont {Carollo}},\ }\bibfield  {title} {\bibinfo {title}
  {Hermitian and non-hermitian topology from photon-mediated interactions},\
  }\href@noop {} {\bibfield  {journal} {\bibinfo  {journal} {Nat. Commun.}\
  }\textbf {\bibinfo {volume} {15}},\ \bibinfo {pages} {2400} (\bibinfo {year}
  {2024})}\BibitemShut {NoStop}%
\bibitem [{\citenamefont {Huang}\ \emph {et~al.}(2024)\citenamefont {Huang},
  \citenamefont {Huang}, \citenamefont {Shen}, \citenamefont {Yves},
  \citenamefont {Pilipchuk}, \citenamefont {Ni}, \citenamefont {Kim},
  \citenamefont {Chiang}, \citenamefont {Powell}, \citenamefont {Zhu} \emph
  {et~al.}}]{huang2024acoustic}%
  \BibitemOpen
  \bibfield  {author} {\bibinfo {author} {\bibfnamefont {L.}~\bibnamefont
  {Huang}}, \bibinfo {author} {\bibfnamefont {S.}~\bibnamefont {Huang}},
  \bibinfo {author} {\bibfnamefont {C.}~\bibnamefont {Shen}}, \bibinfo {author}
  {\bibfnamefont {S.}~\bibnamefont {Yves}}, \bibinfo {author} {\bibfnamefont
  {A.~S.}\ \bibnamefont {Pilipchuk}}, \bibinfo {author} {\bibfnamefont
  {X.}~\bibnamefont {Ni}}, \bibinfo {author} {\bibfnamefont {S.}~\bibnamefont
  {Kim}}, \bibinfo {author} {\bibfnamefont {Y.~K.}\ \bibnamefont {Chiang}},
  \bibinfo {author} {\bibfnamefont {D.~A.}\ \bibnamefont {Powell}}, \bibinfo
  {author} {\bibfnamefont {J.}~\bibnamefont {Zhu}}, \emph {et~al.},\ }\bibfield
   {title} {\bibinfo {title} {Acoustic resonances in non-hermitian open
  systems},\ }\href@noop {} {\bibfield  {journal} {\bibinfo  {journal} {Nat.
  Rev. Phys.}\ }\textbf {\bibinfo {volume} {6}},\ \bibinfo {pages} {11}
  (\bibinfo {year} {2024})}\BibitemShut {NoStop}%
\bibitem [{\citenamefont {Zhang}\ \emph
  {et~al.}(2019{\natexlab{a}})\citenamefont {Zhang}, \citenamefont {Wang},\
  and\ \citenamefont {Song}}]{zhang2019exceptional}%
  \BibitemOpen
  \bibfield  {author} {\bibinfo {author} {\bibfnamefont {K.~L.}\ \bibnamefont
  {Zhang}}, \bibinfo {author} {\bibfnamefont {P.}~\bibnamefont {Wang}},\ and\
  \bibinfo {author} {\bibfnamefont {Z.}~\bibnamefont {Song}},\ }\bibfield
  {title} {\bibinfo {title} {Exceptional-point-induced lasing dynamics in a
  non-hermitian su-schrieffer-heeger model},\ }\href
  {https://doi.org/10.1103/PhysRevA.99.042111} {\bibfield  {journal} {\bibinfo
  {journal} {Phys. Rev. A}\ }\textbf {\bibinfo {volume} {99}},\ \bibinfo
  {pages} {042111} (\bibinfo {year} {2019}{\natexlab{a}})}\BibitemShut
  {NoStop}%
\bibitem [{\citenamefont {Zhang}\ \emph
  {et~al.}(2019{\natexlab{b}})\citenamefont {Zhang}, \citenamefont {Jin},\ and\
  \citenamefont {Song}}]{zhang2019helical}%
  \BibitemOpen
  \bibfield  {author} {\bibinfo {author} {\bibfnamefont {K.~L.}\ \bibnamefont
  {Zhang}}, \bibinfo {author} {\bibfnamefont {L.}~\bibnamefont {Jin}},\ and\
  \bibinfo {author} {\bibfnamefont {Z.}~\bibnamefont {Song}},\ }\bibfield
  {title} {\bibinfo {title} {Helical resonant transport and purified
  amplification at an exceptional point},\ }\href
  {https://doi.org/10.1103/PhysRevB.100.144301} {\bibfield  {journal} {\bibinfo
   {journal} {Phys. Rev. B}\ }\textbf {\bibinfo {volume} {100}},\ \bibinfo
  {pages} {144301} (\bibinfo {year} {2019}{\natexlab{b}})}\BibitemShut
  {NoStop}%
\bibitem [{\citenamefont {Wang}\ \emph
  {et~al.}(2021{\natexlab{a}})\citenamefont {Wang}, \citenamefont {Sweeney},
  \citenamefont {Stone},\ and\ \citenamefont {Yang}}]{wang2021coherent}%
  \BibitemOpen
  \bibfield  {author} {\bibinfo {author} {\bibfnamefont {C.}~\bibnamefont
  {Wang}}, \bibinfo {author} {\bibfnamefont {W.~R.}\ \bibnamefont {Sweeney}},
  \bibinfo {author} {\bibfnamefont {A.~D.}\ \bibnamefont {Stone}},\ and\
  \bibinfo {author} {\bibfnamefont {L.}~\bibnamefont {Yang}},\ }\bibfield
  {title} {\bibinfo {title} {Coherent perfect absorption at an exceptional
  point},\ }\href@noop {} {\bibfield  {journal} {\bibinfo  {journal} {Science}\
  }\textbf {\bibinfo {volume} {373}},\ \bibinfo {pages} {1261} (\bibinfo {year}
  {2021}{\natexlab{a}})}\BibitemShut {NoStop}%
\bibitem [{\citenamefont {Wang}\ \emph
  {et~al.}(2021{\natexlab{b}})\citenamefont {Wang}, \citenamefont {Zhang},\
  and\ \citenamefont {Song}}]{wang2021transition}%
  \BibitemOpen
  \bibfield  {author} {\bibinfo {author} {\bibfnamefont {P.}~\bibnamefont
  {Wang}}, \bibinfo {author} {\bibfnamefont {K.~L.}\ \bibnamefont {Zhang}},\
  and\ \bibinfo {author} {\bibfnamefont {Z.}~\bibnamefont {Song}},\ }\bibfield
  {title} {\bibinfo {title} {Transition from degeneracy to coalescence: Theorem
  and applications},\ }\href {https://doi.org/10.1103/PhysRevB.104.245406}
  {\bibfield  {journal} {\bibinfo  {journal} {Phys. Rev. B}\ }\textbf {\bibinfo
  {volume} {104}},\ \bibinfo {pages} {245406} (\bibinfo {year}
  {2021}{\natexlab{b}})}\BibitemShut {NoStop}%
\bibitem [{\citenamefont {Shi}\ \emph {et~al.}(2022)\citenamefont {Shi},
  \citenamefont {Zhang},\ and\ \citenamefont {Song}}]{shi2022exceptional}%
  \BibitemOpen
  \bibfield  {author} {\bibinfo {author} {\bibfnamefont {Y.}~\bibnamefont
  {Shi}}, \bibinfo {author} {\bibfnamefont {K.}~\bibnamefont {Zhang}},\ and\
  \bibinfo {author} {\bibfnamefont {Z.}~\bibnamefont {Song}},\ }\bibfield
  {title} {\bibinfo {title} {Exceptional spectrum and dynamic magnetization},\
  }\href@noop {} {\bibfield  {journal} {\bibinfo  {journal} {J. Phys.:Condens.
  Matter}\ }\textbf {\bibinfo {volume} {34}},\ \bibinfo {pages} {485401}
  (\bibinfo {year} {2022})}\BibitemShut {NoStop}%
\bibitem [{\citenamefont {Longhi}(2022)}]{longhi2022self}%
  \BibitemOpen
  \bibfield  {author} {\bibinfo {author} {\bibfnamefont {S.}~\bibnamefont
  {Longhi}},\ }\bibfield  {title} {\bibinfo {title} {Self-healing of
  non-hermitian topological skin modes},\ }\href
  {https://doi.org/10.1103/PhysRevLett.128.157601} {\bibfield  {journal}
  {\bibinfo  {journal} {Phys. Rev. Lett.}\ }\textbf {\bibinfo {volume} {128}},\
  \bibinfo {pages} {157601} (\bibinfo {year} {2022})}\BibitemShut {NoStop}%
\bibitem [{\citenamefont {Xue}\ \emph {et~al.}(2025)\citenamefont {Xue},
  \citenamefont {Song}, \citenamefont {Hu},\ and\ \citenamefont
  {Wang}}]{xue2025non}%
  \BibitemOpen
  \bibfield  {author} {\bibinfo {author} {\bibfnamefont {W.-T.}\ \bibnamefont
  {Xue}}, \bibinfo {author} {\bibfnamefont {F.}~\bibnamefont {Song}}, \bibinfo
  {author} {\bibfnamefont {Y.-M.}\ \bibnamefont {Hu}},\ and\ \bibinfo {author}
  {\bibfnamefont {Z.}~\bibnamefont {Wang}},\ }\bibfield  {title} {\bibinfo
  {title} {Non-bloch edge dynamics of non-hermitian lattices},\ }\href@noop {}
  {\bibfield  {journal} {\bibinfo  {journal} {arXiv preprint arXiv:2503.13671}\
  } (\bibinfo {year} {2025})}\BibitemShut {NoStop}%
\bibitem [{\citenamefont {Miao}\ \emph {et~al.}(2025)\citenamefont {Miao},
  \citenamefont {Zhao}, \citenamefont {Wang}, \citenamefont {Qiao},
  \citenamefont {Zhao},\ and\ \citenamefont {Yi}}]{miao2025non}%
  \BibitemOpen
  \bibfield  {author} {\bibinfo {author} {\bibfnamefont {Y.}~\bibnamefont
  {Miao}}, \bibinfo {author} {\bibfnamefont {Y.}~\bibnamefont {Zhao}}, \bibinfo
  {author} {\bibfnamefont {Y.}~\bibnamefont {Wang}}, \bibinfo {author}
  {\bibfnamefont {J.}~\bibnamefont {Qiao}}, \bibinfo {author} {\bibfnamefont
  {X.}~\bibnamefont {Zhao}},\ and\ \bibinfo {author} {\bibfnamefont
  {X.}~\bibnamefont {Yi}},\ }\bibfield  {title} {\bibinfo {title} {Non-abelian
  gauge enhances self-healing for non-hermitian su--schrieffer--heeger chain},\
  }\href@noop {} {\bibfield  {journal} {\bibinfo  {journal} {arXiv preprint
  arXiv:2503.23978}\ } (\bibinfo {year} {2025})}\BibitemShut {NoStop}%
\bibitem [{\citenamefont {Mostafazadeh}(2002)}]{Mostafazadeh2002a}%
  \BibitemOpen
  \bibfield  {author} {\bibinfo {author} {\bibfnamefont {A.}~\bibnamefont
  {Mostafazadeh}},\ }\bibfield  {title} {\bibinfo {title} {Pseudo-hermiticity
  versus pt-symmetry iii: Equivalence of pseudo-hermiticity and the presence of
  antilinear symmetries},\ }\href {https://doi.org/10.1063/1.1489072}
  {\bibfield  {journal} {\bibinfo  {journal} {J. Math. Phys.}\ }\textbf
  {\bibinfo {volume} {43}},\ \bibinfo {pages} {3944} (\bibinfo {year}
  {2002})}\BibitemShut {NoStop}%
\bibitem [{\citenamefont {Su}\ \emph {et~al.}(1979)\citenamefont {Su},
  \citenamefont {Schrieffer},\ and\ \citenamefont {Heeger}}]{Su1979}%
  \BibitemOpen
  \bibfield  {author} {\bibinfo {author} {\bibfnamefont {W.~P.}\ \bibnamefont
  {Su}}, \bibinfo {author} {\bibfnamefont {J.~R.}\ \bibnamefont {Schrieffer}},\
  and\ \bibinfo {author} {\bibfnamefont {A.~J.}\ \bibnamefont {Heeger}},\
  }\bibfield  {title} {\bibinfo {title} {Solitons in polyacetylene},\ }\href
  {https://doi.org/10.1103/PhysRevLett.42.1698} {\bibfield  {journal} {\bibinfo
   {journal} {Phys. Rev. Lett.}\ }\textbf {\bibinfo {volume} {42}},\ \bibinfo
  {pages} {1698} (\bibinfo {year} {1979})}\BibitemShut {NoStop}%
\bibitem [{\citenamefont {Rice}\ and\ \citenamefont
  {Mele}(1982)}]{rice1982elementary}%
  \BibitemOpen
  \bibfield  {author} {\bibinfo {author} {\bibfnamefont {M.~J.}\ \bibnamefont
  {Rice}}\ and\ \bibinfo {author} {\bibfnamefont {E.~J.}\ \bibnamefont
  {Mele}},\ }\bibfield  {title} {\bibinfo {title} {Elementary excitations of a
  linearly conjugated diatomic polymer},\ }\href
  {https://doi.org/10.1103/PhysRevLett.49.1455} {\bibfield  {journal} {\bibinfo
   {journal} {Phys. Rev. Lett.}\ }\textbf {\bibinfo {volume} {49}},\ \bibinfo
  {pages} {1455} (\bibinfo {year} {1982})}\BibitemShut {NoStop}%
\end{thebibliography}

\end{document}